\documentstyle[12pt]{article}

\input mssymb

 1
\font\subscriptsizebbfont=msym7 scaled \magstep 1
 1
 3
\font\bbfont=msym10 scaled \magstep1  

\def\subscriptsizeBbb#1{\hbox{\subscriptsizebbfont #1}}

\def\Bbb#1{\hbox{\bbfont #1}}

\input psfig
\psfiginit

\begin{document}

\begin{titlepage}

$ $

\vspace{-2.5cm}

\noindent\hspace{-.5cm}
\parbox{4cm}{September 1996}\
  \hspace{9cm}\
  \parbox{5cm}{{\sc umtg} -- {\small 191}   \newline
  {\tt hep-th/9610042}  }

\vspace{1.8cm}

\centerline{\large\bf Topological Membrane Solitons and}
\vspace{.2cm}
\centerline{\large\bf Loop Orders of Membrane Scatterings}
\vspace{.2cm}
\centerline{\large\bf in M-theory}

\vspace{1.5cm}

\centerline{\large
 Chien-Hao Liu\footnote{e-mail: chienliu@phyvax.ir.miami.edu}}

\vspace{1em}

\centerline{\it Department of Physics}
\centerline{\it University of Miami}
\centerline{\it P.O. Box 248046}
\centerline{\it Coral Gables, FL.\ 33124}

\vspace{.8cm}

\begin{quotation}
\centerline{\bf Abstract}

\vspace{0.3cm}

\baselineskip 13pt

{\small Two topological issues on membranes in M-theory are studied:
 (1) Soliton is an important subject in M-theory. Under the framework
     of obstruction theory with the help from framed links in $S^3$,
     we give a complete enumeration of topological membrane solitons in
     a string-admissible target-space of the form a product of
     Minkowskian space-times, tori, and K3-surfaces. Patching of these
     solitons and their topological charges are also defined and
     discussed.
 (2) Loop order of membrane scatterings is the basis for a perturbative
     M-theory. We explore this concept with emphases on its distinct
     features from pointlike and stringlike particles.
 For completeness, a light exposition on homologies of compact oriented
 $3$-manifolds is given in the Appendix.
} 
\end{quotation}

\bigskip

\baselineskip 12pt


\bigskip

\noindent
MSC number 1991: 55Q05, 83E30; 55N25, 57M25, 57N10, 81T30.

\bigskip

\baselineskip 12pt

{\footnotesize
\noindent{\bf Acknowledgements.}
I would like to thank Orlando Alvarez, Hung-Wen Chang, Michael Duff,
 Brian Greene, Emil Martinec, Rafael Nepomechie, and William Thurston
 for invaluable inspirations, discussions, help, and suggestions in the
 preparation of this paper.
Much of the work is done at
 TASI-96 (U.Colorado-Boulder, by B.G. and Kalyana Mahanthappa),
 Duality-96 (Argonne, by Thomas Curtright and Cosmas Zachos),
 and U.C.\ Berkeley. I would like to thank these institutions,
 organizers, and participants.
In particular, I am greatly indebted to Paul Aspinwall, M.D., B.G.,
 Jeffrey Harvey, David Morrison, and Hirosi Ooguri for their series of
 lectures; and Enrique Alvarez, Ctirad Klim\u{c}\'{i}k, Yolanda Lozano,
 H.O., and Joseph Polchinski for conversations.
I also like to thank Yi-Chun Chou and Ta-Chun Yao for hospitality and
 Ling-Miao Chou for encouragement.
Unavoidably, I recall the joy, pain, and efforts to understand
 Thurston's work with Noel Brady, Bill Grosso, and Inkang Kim for
 nearly two years at U.C.B.. My deepest thanks to them.
} 

\end{titlepage}

\baselineskip 15pt

\newpage
\pagenumbering{arabic}

\begin{flushleft}
{\Large\bf 0. Introduction and outline.}
\end{flushleft}
\vspace{-3cm}

\centerline{\sc Membrane Solitons and Loop Orders in M-Theory}

\vspace{2cm}

\begin{flushleft}
{\bf Introduction.}
\end{flushleft}
The string $\beta$-function at 1-loop level requires that the string
target-space $N$ be Ricci flat. Such string-admissible target-spaces
include products of the Minkowskian space-time
${\Bbb R}^{l+1}\times [0,1]^{l^{\prime}}$, where $[0,1]$ is the
quotient orbifold of $S^1$ by an orientation-reversing diffeomorphism,
$n$-dimensional tori ${\Bbb T}^n$, K3-surfaces, and Calabi-Yau
$3$-manifolds. Depending on the fermionic fields and supersymmetries
involved in the theory, there are several types of strings
(cf.\ [G-S-W]). Recently, the progress in our understanding of
dualities indicates that these different types of strings may be
viewed miraculously as "boundary values" or different "limits" of a
master (or mother) theory, {\it M-theory} at an $11$-dimensional
string-admissible target-space. (Readers are referred to, for example,
the physics literatures listed in Reference, which motivate and
influence this paper.)

Exactly what M-theory is remains a mystery. Besides its relation
to strings, it has $11$-dimensional supergravity as a low energy limit.
The theory contains not only membranes but also other higher
dimensional $p$-branes moving around in a string-admissible
target-space. Due to our limited knowledge of higher dimensional
manifolds and general features of Calabi-Yau $3$-manifolds as
CW-complexes, we confine ourselves in this paper only to membranes
moving in a string-admissible target-space $N$ of the form
$$
 N\; =\;
   ({\Bbb E}^{l+1}\times [0,1]^{l^{\prime}})\times {\Bbb T}^n\times
     \mbox{\rm K3-surface}\,;
$$
and consider two topological issues in M-theory:
(1) topological membrane solitons in $N$, and
(2) loop orders of membrane scatterings.

Just as the non-vanishness of homotopy groups provides the topological
reason some solitons in gauge theory appear, the fact that there are
non-trivial homotopy classes of maps from a membrane world-volume $M^3$
into $N$ can also provide the topological reason for the appearance of
some membrane solitons in M-theory. Each homotopy class may represent a
family of physical solitons that can be continuously deformed into each
other and thus shall be regarded as a {\it topological soliton}.
The set $[M^3,N]$ of all of them should then play the same role to
membrane soliton problems in M-theory as homotopy groups to soliton
problems in gauge theory (cf.\ [Co]). Under the framework of
obstruction theory with the help from framed links in $S^3$,
we give a complete enumeration of the classes in $[M^3,N]$.
Their patching and charges are also defined and discusssed.

Next we consider another topological issue, the loop order of
membrane scatterings. In string theory, the loop order of a world-sheet
that describes a family of string scattering processes is a
well-defined concept and it plays the role as an expansion parameter for
the perturbative string theory. We shall find that, once going beyond
strings to higher dimensional extended objects, some essential features
become very different and work remains to be done to fix this concept.

\bigskip

\begin{flushleft}
{\bf Outline.}
\end{flushleft}
{\small

\baselineskip 13pt

1. Essential mathematical backgrounds.

2. Enumeration of topological membrane solitons in M-theory.

3. Patching of membrane solitons and topological charges.

4. Loop orders of membrane scatterings.

Appendix: A light exposition on homologies of $3$-manifolds.
} 

\vspace{.3cm}

{\small
\begin{quote}
 {\it "I have merely transmitted what was taught to me $\cdots\cdots$."}
    \newline \rule{0ex}{3ex}
    \parbox[t]{13cm}{\raggedleft --------- \hspace{1em}
                                      Confucius, 551? -- 479 B.C.}
\end{quote}
} 

\vspace{.3cm}

\baselineskip 15pt

\section{Essential mathematical backgrounds.}

For the convenience of physicists and the introduction of notations,
we sketch in this section some mathematical preliminaries needed for
this paper. Some are given only in key words and more details are
referred to the literatures.

\bigskip

\noindent $\bullet$
{\bf Algebraic and differential topology.}
([B-T], [D-F-N], [Mi3], [Sp2], and [Vi].)
CW-complexes $X$; their homologies $H_{\ast}(X;{\Bbb Z})$ and
 cohomologies $H^{\ast}(X;{\Bbb Z})$;
homotopy groups $\pi_{\ast}(X,\cdot)$;
Eilenberg-MacLane spaces $K(\pi,n)$ as classifying spaces for
 cohomologies;
K\"{u}nneth formula;
the universal coefficient theorem;
homotopy classes of maps between two complexes;
Hopf-Whitney classification theorem [Wh];
de Rham cohomology $H_{DR}^{\ast}(N)$ for smooth manifold $N$;
relative de Rham cohomology $H_{DR}^{\ast}(N,N_0)$, where $N_0$ is
a submanifold of $N$;
Sard's theorem;
Pontryagin-Thom construction.

When the base-point does not play roles in the discussion,
we shall frequently omit it and denote the homotopy groups of $X$
simply by $\pi_{\ast}(X)$.

\bigskip

\noindent $\bullet$
{\bf Heegaard splitting.} ([He], [Ja], [Si] and [Sti].)
A {\it handlebody $H_g$ of genus $g$} is a $3$-ball $B^3$ with $2g$
disjoint $2$-discs in its boundary glued in pairs by an
orientation-reversing homeomorphism. It has boundary a surface
$\Sigma_g$ of genus $g$. Every closed oriented $3$-manifold $M^3$ can be
obtained by gluing two handlebodies of the same genus along their
boundary. Such decomposition is called a {\it Heegaard splitting} of
$M^3$. The data of the splitting can be coded by $\Sigma_g$
together with two systems of simple loops $\{C_1,\cdots,C_g\}$ and
$\{C_1^{\prime},\cdots,C_g^{\prime}\}$ with each system cutting
$\Sigma_g$ into a $2$-sphere with $g$ pairs of holes. {}From a
Heegaard splitting, one has a special CW-complex structure of $M^3$
constructed by first removing from $\Sigma_g$ a $2$-disc disjoint
with all the loops to obtain a surface-with-one-hole $\Sigma_g^{\circ}$,
next attaching a $2$-cell $e^2_i$ (and resp.\ $e^{2\,\prime}_i$) along
$C_i$ (resp.\ $C_i^{\prime}$) to obtain a $2$-complex $X^2$, and finally
attaching a $3$-cell $e^3$ to $X^2$ as indicated in {\sc Figure 1.1}.
We shall call this a
{\it CW-complex structure of $M^3$ associated to a Heegaard splitting}.
The structure $M^3=X^2\cup_h e^3$, where $h$ is the attaching map, shall
be important to us later. Up to homotopy, we shall assume that $h$ is
an immersion on the complement of a $1$-dimensional subset in
$\partial\,e^3$ (cf.\ {\sc Figures 1.1} and {\sc 2.1}).

\begin{figure}[htbp]
\setcaption{{\sc Figure 1.1.}  A CW-complex structure of $M^3$
    associated to a Heegaard splitting. This is an example of $g=2$.
    For clarity, we duplicate the $\Sigma_g^{\circ}$-part of $X^2$ in
    the picture, with one having $e^2_i$ attached and the other having
    $e^{2\,\prime}_i$ attached. }
\centerline{\psfig{figure=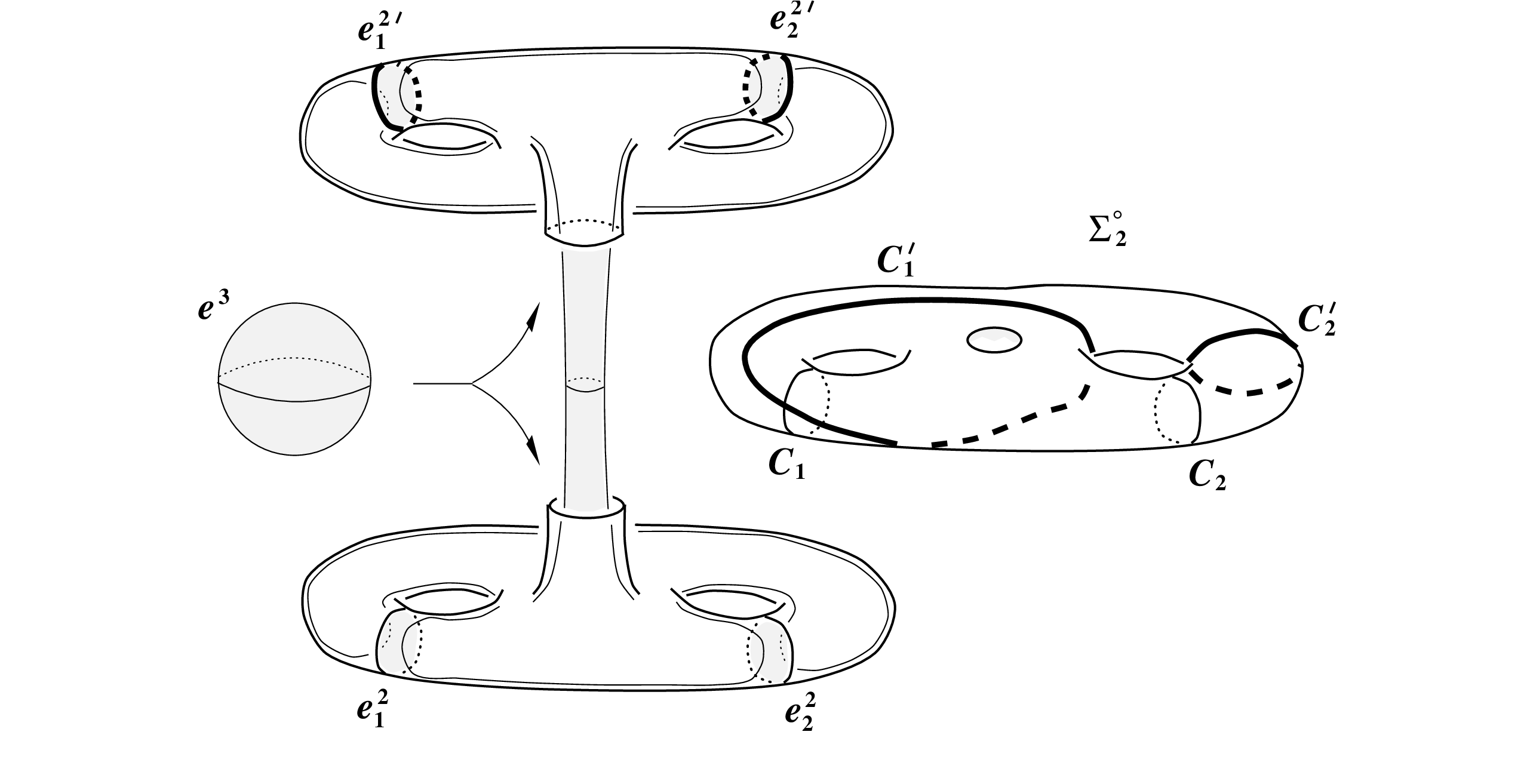,width=13cm,caption=}}
\end{figure}

For this complex, recall the exact sequence
$$
 H_3(M^3,X^2;{\Bbb Z})\;\stackrel{\partial}{\longrightarrow}\;
   H_2(X^2;{\Bbb Z})\;\longrightarrow\; H_2(M^3;{\Bbb Z})\;
   \longrightarrow\; 0\,.
$$
Since $H_3(M^3,X^2;{\Bbb Z})$ is generated by $e^3$ with
$\partial\,e^3 \sim 0$ in $H_2(X^2;{\Bbb Z})$, the inclusion map from
$X^2$ into $M^3$ induces an isomorphism from $H_2(X^2;{\Bbb Z})$ onto
$H_2(M^3;{\Bbb Z})$.
(Consequently, $H_2(M^3;{\Bbb Z})$ is torsion free.)

Suppose that $M^3$ is the connected sum $M^3_1\,\sharp\,M^3_2$ of two
closed orientable $3$-manifolds, $M^3_1$, $M^3_2$. Let
$M^3_i=X^2_i\cup_{h_i}\,e^3_i$, $i=1,\,2$, from some Heegaard
splittings. One can get a compatible complex structure
$M^3=X^2\cup_h\,e^3$ by first band-connected-summing $X^2_1$, $X^2_2$
along $\partial\Sigma_{g_1}^{\circ}$ and $\partial\Sigma_{g_2}^{\circ}$
to obtain a $2$-complex $X^2$, and then constructing $(e^3,h)$ by
amalgamating $(e^3_1, h_1)$ and $(e^3_2, h_2)$ together, following the
pasting when connected-summing $M^3_1$ and $M^3_2$
({\sc Figure 1.2}, cf.\ [Ja]). Recall that
$H_i(M^3;{\Bbb Z})=H_i(M^3_1;{\Bbb Z})\oplus H_i(M^3_2;{\Bbb Z})$,
for $i=1$, $2$.

\begin{figure}[htbp]
\setcaption{{\sc Figure 1.2.} Connected-summing $M^3_1$ and $M^3_2$
    along an embedded $S^2$ in $M^3_1$ that bounds a $3$-ball $B^3$
    which intersects $\Sigma_{g_1}^{\circ}$ with a half-$2$-disc along
    $\partial\Sigma_{g_1}^{\circ}$ and a similar $S^2$ in $M^3_2$.
    The relation between $e^3_1$, $e^3_2$ (with part of their boundary
    pasted along $\partial\Sigma_{g_1}^{\circ}$ and
    $\partial\Sigma_{g_2}^{\circ}$ respectively) and $e^3$ are
    indicated. Each of the thickened-half-shells $S_{1,1}$, $S_{1,2}$,
    $S_{2,1}$, and $S_{2,2}$ forms a quarter of $e^3$ after pasting
    their inner boundary following the operation of connected sum.
    The unpasted outer boundaries become the four spots on
    $\partial\,e^3$. They are to be attached to
    $\Sigma_{g_1+g_2}^{\circ}$ following their memories of the old
    attaching maps from $M^3_1$ and $M^3_2$.
    Note that, after being pasted, the shaded part of $\partial\,e^3$
    forms a collar of $\partial\Sigma_{g_1+g_2}^{\circ}$ in
    $\Sigma_{g_1+g_2}^{\circ}$. }
\centerline{
   \psfig{figure=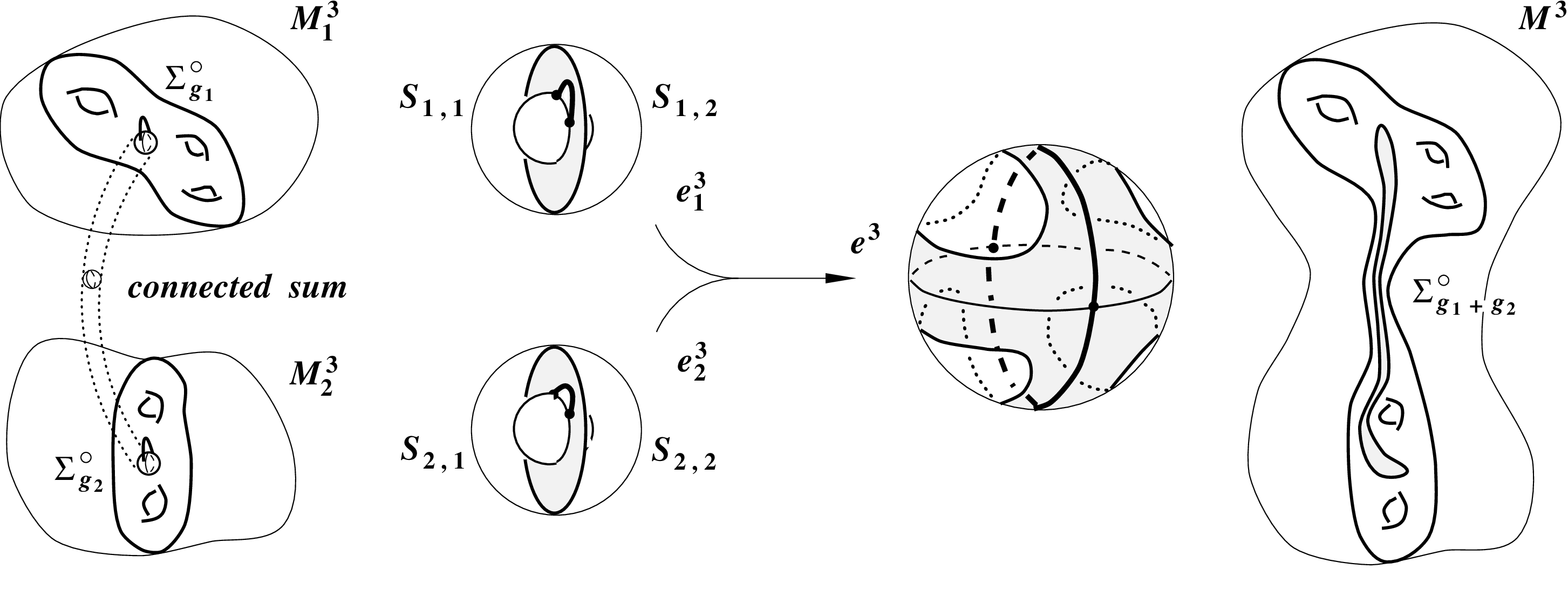,width=13cm,caption=}}
\end{figure}

\bigskip

\noindent $\bullet$
{\bf Bouquets and framed links.}
([D-K], [Hi], [Ki], and [Mi3].)
A {\it bouquet of $2$-spheres}, denoted by $\vee_r\,S^2$, is a union of
$2$-spheres with a single common point (the {\it vertex} $\ast$). Its
homotopy groups have been studied by Hilton in [Hi]. We shall need the
first three:
\begin{quote}
 \hspace{-2em} (1)
  $\pi_1(\vee_r\,S^2)\,=\,0$.

 \hspace{-2em} (2)
  $\pi_2(\vee_r\,S^2)\;=\;\oplus_r\,{\Bbb Z}$
  is generated by the component $S^2$'s of the bouquet.

 \hspace{-2em} (3)
  $$
   \pi_3(\vee_r\,S^2)\;
    =\;(\oplus_r\,{\Bbb Z})\; \mbox{$\bigoplus$}\;
         (\oplus_{(\hspace{-.8ex}\mbox{\tiny $\begin{array}{c}
               r \\ 2 \end{array}$}\hspace{-.8ex})}\,{\Bbb Z}) \;
    =\; \oplus_{\frac{r(r+1)}{2}}\,{\Bbb Z}
  $$
  is generated by two collections of elements: The generators $[i]$,
  $i=1,\cdots, r$, of the first summand correspond to maps from $S^3$
  to some component $S^2$ that give the Hopf fibration of $S^3$. As an
  attaching map of a $4$-cell, each gives a $\mbox{{\Bbb C}P}^2$. The
  generators $[i,j]$, $1\leq i<j\leq r$, of the second summand
  correspond to maps from $S^3$ to some sub-bouquet $S^2\vee S^2$,
  each of which, as an attaching map of a $4$-cell, gives an
  $S^2\times S^2$.
\end{quote}

There is a geometric interpretation for $\pi_3(\vee_r\,S^2)$.
Given a map $f$ from an oriented $S^3$ to an oriented $\vee_r\,S^2$.
One may assume that $f$ is smooth on the open subset of $S^3$ that is
mapped to the complement of the vertex $\ast$. By Sard's theorem, there
exist regular values $y_1,\cdots,y_r$ of $f$, one on each oriented
$S^2$ in $\vee_r\,S^2$. The preimage $f^{-1}(y_i)$ is an oriented framed
link in $S^3$. For example, the framed links and their linking matrix
associated to the generators $[i]$ and $[i,j]$ are indicated in
{\sc Figure 1.3}.
\begin{figure}[htbp]
\setcaption{{\sc Figure 1.3.} The oriented framed link and the linking
    matrix associated to the generators $[i]$ and $[i,j]$ of
    $\pi_3(\vee_r S^2)$. The framing is indicated by a band along each
    component knot; and all the missing entries of the matrix are zero.}
\centerline{\psfig{figure=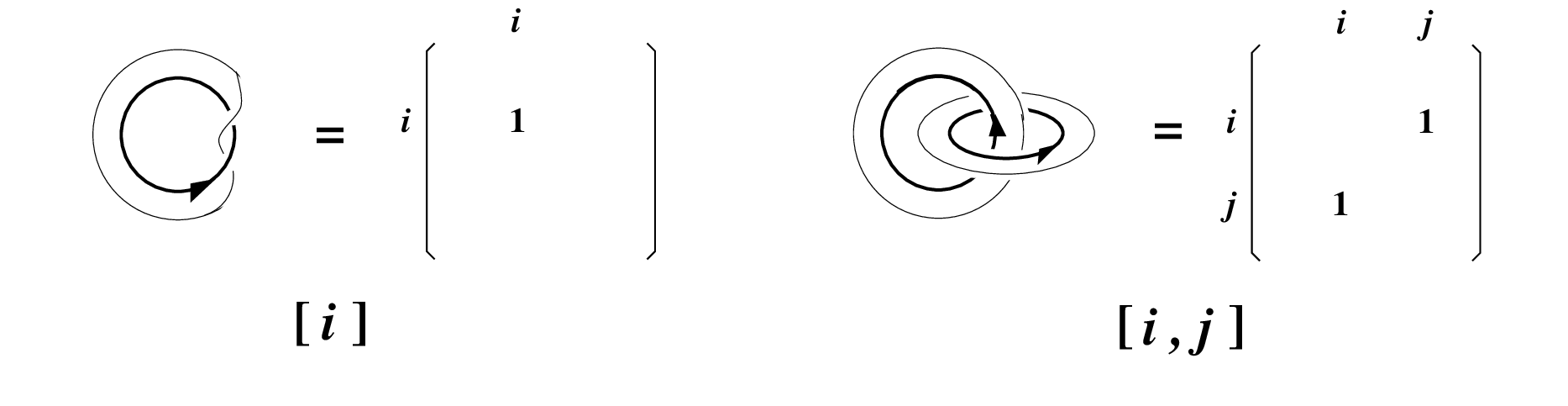,width=13cm,caption=}}
\end{figure}
Two maps $f_1$ and $f_2$ from $S^3$ to $\vee_r\,S^2$ are homotopic if
and only if their associated collections of framed links have the same
linking matrix. On the other hand, any symmetric $r\times r$ integral
matrix can be realized as the linking matrix associated to some $f$ by
the Pontryagin-Thom construction. Thus one has a natural isomorphism
between $\pi_3(\vee_r\,S^2)$ and the additive group of symmetric
$r\times r$ integral matrices. Explicitly,
$$
  \sum_i\,c_i\,[i]\,+\,\sum_{i<j}\,c_{ij}\,[i,j]
      \hspace{2em} \longleftrightarrow \hspace{2em}
  \left( \begin{array}{llll}
            c_1    & c_{12} & \cdots & c_{1r} \\
            c_{12} & c_2    & \cdots & c_{2r} \\
            \vdots & \vdots & \vdots & \vdots \\
            c_{1r} & c_{2r} & \cdots & c_r
         \end{array} \right)\;\;.
$$

\bigskip

\noindent $\bullet$
{\bf Four-manifolds.}
([D-K], [Mi1], [Ki], and [Whi].)
A compact simply-connected orientable $4$-manifold is homotopic to
a bouquet of $2$-spheres attached with a $4$-ball. In notation,
$M^4\;=\;(\vee_r\,S^2) \cup_h\, e^4$, where $h$ is an attaching map.
With respect to the basis consisting of the component $S^2$'s of the
bouquet, the intersection form of $H_2(M^4;{\Bbb Z})$ coincides with the
linking matrix of the oriented framed links in $S^3$ associated to $h$.

\bigskip

\noindent $\bullet$
{\bf K3-surfaces.}
([D-K], [Ki], [Mi1], and [Sp1].)
K3-surfaces are compact complex $2$-manifolds with vanishing first Chern
class. These Calabi-Yau 2-folds are all diffeomorphic to each other,
whose topology can be constructed as follows. Observe that the
$\{\pm 1\}$-action on a complex $4$-torus ${\Bbb T}^4$ by
multiplication has $2^4=16$ fixed points. This action extends to one on
the blow-up ${\Bbb T}^4\,\sharp\, 16\,\mbox{{\Bbb C}{\rm P}}^2$ of
${\Bbb T}^4$ at these fixed points. The latter action leaves the
exceptional divisor - $16\, \mbox{{\Bbb C}{\rm P}}^1$'s - fixed. The
quotient is then an oriented 4-manifold K3 that supports
all K3-surfaces. Being simply-connected, its oriented homotopy type is
determined by its intersection form
$$
 3\,\left( \begin{array}{cc}
                 0 & 1 \\
                 1 & 0
           \end{array}      \right) \oplus 2\,(-E_8)
$$
on $H_2(\mbox{\rm K3}\,;{\Bbb Z})$, where $E_8$ is the positive-definite
even form of rank $8$ given by the matrix: (Only non-zero entries are
shown.)
$$
 \left(    \mbox{\small
 $\begin{array}{rrrrrrrr}
       2 &    & -1 &   & & & & \\
         &  2 &    & -1 & & & & \\
      -1 &    &  2 & -1 & & & & \\
         & -1 & -1 &  2 & -1 &&& \\
         &    &    & -1 &  2 & -1 && \\
         &    &    &    & -1 &  2 & -1 & \\
         &    &    &    &    & -1 &  2 & -1 \\
         &    &    &    &    &    & -1 &  2
 \end{array}$  }
 \right)\,.
$$
Up to homotopy, K3 is a bouquet of $22$ oriented $2$-spheres
attached with a $4$-ball $e^4$. In notation,
$$
 \mbox{\rm K3}\,= (\vee_{22}\,S^2)\cup_{h_0} e^4
$$
for some attaching map $h_0$.

{}From the previous Items and Hurewicz isomorphism, one has
$$
 \pi_2(\mbox{\rm K3})\;=\;H_2(\mbox{\rm K3};{\Bbb Z})\;=\;
   \oplus_{22}\,{\Bbb Z}
$$
and
$$
 \pi_3(\mbox{\rm K3})\;
   =\;\left.\mbox{\raisebox{1ex}{$\pi_3(\vee_{22}\,S^2)$}} \right/
            \mbox{\raisebox{-1ex}{$[h_0]\sim 0$}}  \;\;\;,
$$
where $[h_0]$ is the class in $\pi_3(\vee_{22}\,S^2)$ represented by
$h_0$. Explicitly,
\begin{eqnarray*}
 \lefteqn{ [h_0]\;=\; [1,2]\,+\,[3,4]\,+\,[5,6]\,
                                   -\, 2\,\sum_{i=7}^{22}\,[i]} \\
          & & +\, [7,9]\,+\,[8,10]\,+\,[9,10]\,+\,[10,11]\,
                    +\,[11,12]\,+\,[12,13]\,+\,[13,14]     \\
          & & +\, [15,17]\,+\,[16,18]\,+\,[17,18]\,+\,[18,19]\,
                    +\, [19,20]\,+\,[20,21]\,+\,[21,22]
\end{eqnarray*}
({\sc Figure 1.4}). Since the coefficient of, say, $[1,2]$-component
in the combination is $1$, one can replace $[1,2]$ by $[h_0]$ to form
a new basis of $\pi_3(\vee_{22}\,S^2)$ with the rest of $[i]$'s and
$[i,j]$'s. Consequently, after modding out $[h_0]$, one has
$$
 \pi_3(\mbox{\rm K3})\; =\; \oplus_{252}\,{\Bbb Z}\,.
$$

\begin{figure}[htbp]
\setcaption{{\sc Figure 1.4.} The intersection matrix of K3,
    coded in a Dynkin diagram, gives $[h_0]$. The dotted vertex
    indicates self-intersection number $0$ and the undotted one $-2$.
    Each edge indicates intersection number $1$.}
\centerline{\psfig{figure=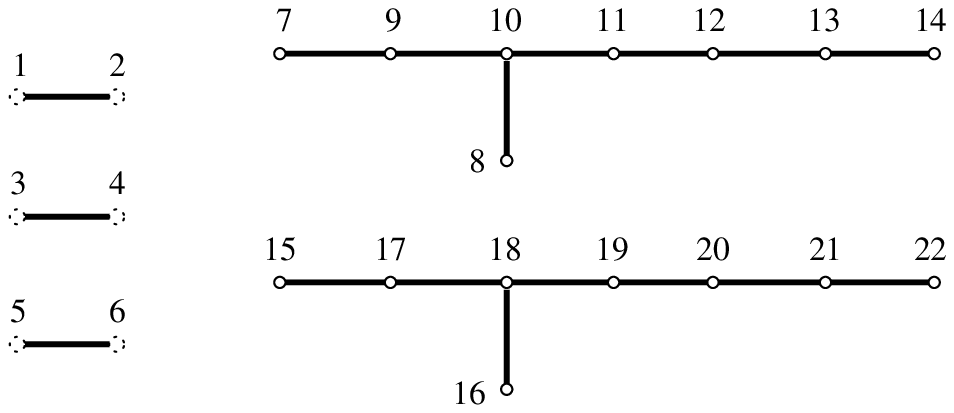,width=13cm,caption=}}
\end{figure}

\bigskip

\noindent $\bullet$
{\bf Obstruction theory and enumeration of map-classes.}
([Hu], [Sp2], and [Wh].)
Given two CW-complexes $X$ and $Y$. A {\it map-class} from $X$ to $Y$
is a homotopy class of maps from $X$ to $Y$. They form a set $[X,Y]$
that depends only on the homotopy type of the source and target
complexes. Let $X^{(n)}$ be the $n$-skeleton of $X$. Two maps
$f_1$, $f_2$ from $X$ to $Y$ are said to be {\it $n$-homotopic} if the
restrictions $f_1|_{X^{(n)}}$, $f_2|_{X^{(n)}}$ are homotopic. This is
a topological concept, independent of what CW-complex structures are
used for $X$ and $Y$ in the definition. An $n$-homotopy class
represented by $f$ shall be denoted by $[f]_n$ and the set of all such
classes by $[X,Y]_n$.

Assume that $Y$ is path- and simply-connected.
Let $w$ be an $n$-homotopy class of maps from $X$ to $Y$ and
$\mbox{\it Map}\,(X^{(n)},Y)$ be the space of all maps from $X^{(n)}$
to $Y$ (with the compact-open topology). Fix an $f_0$ in $w$ and define
$J_w$ to be the image of the following homomorphism:
$$
\begin{array}{cccl}
 \pi_1(\mbox{\it Map}\,(X^{(n)},Y)\,,\,f_0|_{X^{(n)}})
     & \longrightarrow
         & H^{n+1}(X;\pi_{n+1}(Y))  & \\[1ex]
   [h_t]  & \longmapsto  & \delta^{n+1}(f_0,f_0;h_t) &,
\end{array}
$$
where $h_t$ is a homotopy of $f_0|_{X^{(n)}}$ to itself,
and $\delta^{n+1}(f_0,f_0;h_t)$ is the
{\it obstruction cohomology class} in $H^{n+1}(X;\pi_{n+1}(Y))$
determined by $f_0$ and $h_t$. (One may think of $J_w$ as consists of
all the "fake obstructions" for two $n$-homotopic $f_1$, $f_2$ in $w$
to be $(n+1)$-homotopic.) As the notation already indicates, $J_w$
depends only on $w$. The set of $(n+1)$-homotopy classes of maps
in $w$ can then be described by a subset $A_{f_0}$ in the quotient
$H^{n+1}(X;\pi_{n+1}(Y))/J_w$, which consists of all the $J_w$-orbit
of obstruction classes $\delta^{n+1}(f_0,f;h_t)$ in
$H^{n+1}(X;\pi_{n+1}(Y))$, where $f$ is in $w$ and $h_t$ is
an $n$-homotopy between $f_0$ and $f$. For different choices of $f_0$
in $w$, $A_{f_0}$ differ by a translation in
$H^{n+1}(X;\pi_{n+1}(Y))/J_w$.

\bigskip

\section{Enumeration of topological membrane solitons in M-theory.}

Given a compact oriented $3$-manifold $M^3$ that describes a
membrane world-volume. Let $N$ be the string-admissible target-space
in the product form
$$
 N\;=\; {\Bbb R}^{l+1}\times [0,1]^{l^{\prime}}
          \times {\Bbb T}^n \times \mbox{\rm K3}\,.
$$
As explained and defined in the Introduction, the set of topological
membrane solitons supported by $M^3$ is then the set $[M^3,N]$ of
map-classes from $M^3$ to $N$. Since
$$
 [M^3,N]\;=\;[M^3,{\Bbb R}^{l+1}\times [0,1]^{l^{\prime}}]\,
   \times\, [M^3,{\Bbb T}^n]\, \times\, [M^3,\mbox{\rm K3}]\,,
$$
we only need to understand $[M^3,\,\cdot\,]$ for each component of
the product.

\bigskip

\noindent
{\it Remark 2.1.} When $\partial M^3$ is non-empty, $[M^3,N]$
 enumerates topological membrane solitons, supported by $M^3$, with
 free boundary. Let $\partial_0 M^3$ be the union of some components
 of $\partial M^3$. Then one may also consider the set
 $[(M^3, \partial_0 M^3), (N,\partial N)]$ of map-classes from
 $M^3$ to $N$ with $\partial_0 M^3$ mapped to $\partial N$. They
 correspond to membrane solitons with part of the boundary components
 confined in the boundary of $N$. For the $N$ in our problem, there is
 a natural surjection
 $$
  [\,(M^3, \partial_0 M^3)\,,\, (N,\partial N)\,]\;
     \longrightarrow\; [M^3,N]\,,
 $$
 the preimage of which at each point is isomorphic to
 $[\partial_0 M^3, S^{l^{\prime}-1}]$. Explicitly,
 \begin{quote}
  \parbox[t]{3em}{$l^{\prime}=1$} \hspace{1em}
   \parbox[t]{11cm}{
    $[\partial_0 M^3,\{0,1\}]\,
          =\,2^{\pi_0(\partial_0 M^3)}$,
    where $\pi_0(\partial_0 M^3)$ is the set of components of
    $\partial_0 M^3$.                }

  \parbox[t]{3em}{$l^{\prime}=2$} \hspace{1em}
   \parbox[t]{11cm}{
    $[\partial_0 M^3, S^1]\,=\, H^1(\partial_0 M^3;{\Bbb Z})$.   }

  \parbox[t]{3em}{$l^{\prime}=3$} \hspace{1em}
   \parbox[t]{11cm}{
    $[\partial_0 M^3, S^2]\,
     =\,\oplus_{\pi_0(\partial_0 M^3)}{\Bbb Z}$, following from the
    fact that the homotopy class of a map from a connected oriented
    closed surface to $S^2$ is determined by its degree. }

  \parbox[t]{3em}{$l^{\prime}\geq 4$} \hspace{1em}
   \parbox[t]{11cm}{
    $[\partial_0 M^3, S^{l^{\prime}-1}]$ is a singleton; and hence
    $[(M^3,\partial_0 M^3),(N,\partial N)]$ and $[M^3,N]$ are
    canonically isomorphic. }
 \end{quote}
 Consequently, one only needs to understand $[M^3, N]$; and all other
 boundary-confined situations then follow.

\bigskip

\begin{flushleft}
{\bf Minkowskian and toroidal target-spaces.}
\end{flushleft}
The Minkowskian space-time ${\Bbb R}^{l+1}\times [0,1]^{l^{\prime}}$
is contractible and hence contributes no obstructions to deformations
of maps into it.

The circle $S^1$ is an Eilenberg-MacLane space $K({\Bbb Z},1)$.
Thus the map-classes from $M^3$ into it are completely classified
by $H^1(M^3;{\Bbb Z})$. Consequently,
$$
 [M^3,{\Bbb T}^n]\;=\; \prod_n\,[M^3,S^1]\;
   =\; \prod_n\,H^1(M^3;{\Bbb Z}) \;=\;
            \left(\oplus_n{\Bbb Z}\right)^{b_1}\,,
$$
where $b_1$ is the first Betti number of $M^3$.

\bigskip

\begin{flushleft}
{\bf K3 target-spaces: (1) When $\partial\, M^3$ is non-empty.}
\end{flushleft}
Since in this case $M^3$ is homotopic to a 2-complex $X^2$ and K3 is
simply-connected, by the Hopf-Whitney classification theorem the
map-classes from $M^3$ to K3 are completely classified by elements in
$H^2(X^2;\pi_2(\mbox{\rm K3}))=H^2(M^3;\pi_2(\mbox{\rm K3}))$. By the
universal coefficient theorem and the fact that, for any finite
abelian group $A$, the group of extensions $\mbox{\it Ext}\,(A,B)$
is isomorphic to $A\otimes B$ for any abelian group $B$, one has
$$
 [M^3,\mbox{\rm K3}]\;=\;H^2(M^3;\pi_2(\mbox{\rm K3}))\;
   =\; (\oplus_{22}{\Bbb Z})^{b_2}\,\mbox{$\bigoplus$}\,
     \left( \oplus_{22}\,\mbox{\it Tor}\,H_1(M^3;{\Bbb Z})\right)\,,
$$
where $b_2$ is the second Betti number of $M^3$ and
$\mbox{\it Tor}\,H_1(M^3;{\Bbb Z})$ is the torsion part of
$H_1(M^3;{\Bbb Z})$.

\bigskip

\begin{flushleft}
{\bf K3 target-spaces: (2) When $M^3$ is closed.}
\end{flushleft}
Obstruction theory explains how maps may not be deformed into each
other and hence different map-classes can arise, but in general it
alone is not powerful enough to enumerate all the map-classes. In the
current case, however, it turns out that, while under the frame work
of obstruction theory, one can relate the problem to framed links in
$S^3$ and hence resolves the difficulties.

\bigskip

\noindent
{\it (i) Preparations.}
Given a Heegaard splitting of $M^3$ along a closed oriented surface
$\Sigma_g$ with two systems of characteristic loops
$\{C_1, \cdots, C_g\}$ and $\{C_1^{\prime},\cdots,C_g^{\prime}\}$. Let
$M^3=X^2\cup_h e^3$ be a CW-complex structure of $M^3$ associated to
the splitting. The attaching map $h$ from $\partial\,e^3=S^2$ to $X^2$
induces a decomposition of $S^2$: first, the loop
$h^{-1}(\partial\Sigma_g^{\circ})$ separates $S^2$ into the union of
two hemispheres $D^2_-$ and $D^2_+$; and then the connected components
of $h^{-1}(C_i)$, $h^{-1}(C_i^{\prime})$, and
$h^{-1}(X^2-\cup_iC_i -\cup_i C_i^{\prime})$ gives a further
decomposition of each hemisphere. We shall denote the $2$-dimensional
components in $D^2_-$ (resp.\ $D^2_+$) by $\Omega_j$ (resp.\
$\Omega_j^{\prime}$) for those from the $\Sigma_g^{\circ}$ part and
$e^2_{i,1}$, $e^2_{i,2}$ (resp.\ $e^{2\,\prime}_{i,1}$,
$e^{2\,\prime}_{i,2}$) for those from the $e^2_i$
(resp.\ $e^{2\,\prime}_i$) part. By construction, there is an
orientation-reversing homeomorphism $\rho$ from $\Omega_j$ to
$\Omega_j^{\prime}$ and from $e^2_{i,1}$ (resp.\ $e^{2\,\prime}_{i,1}$)
to $e^2_{i,2}$ (resp.\ $e^{2\,\prime}_{i,2}$) by sending $p$ to
$\hat{p}$ if $p$ and $\hat{p}$ are attached to the same point on $X^2$.
As a map on $S^2$, $\rho$ is in general discontinuous.
({\sc Figures 1.1} and {\sc 2.1.})

\begin{figure}[htbp]
\setcaption{{\sc Figure 2.1.} Two examples of the decomposition of
   $S^2$ induced from Heegaard splittings. In each example, the map
   $\rho$ is indicated by its effect on embedded letters K and R.}
\centerline{
 \psfig{figure=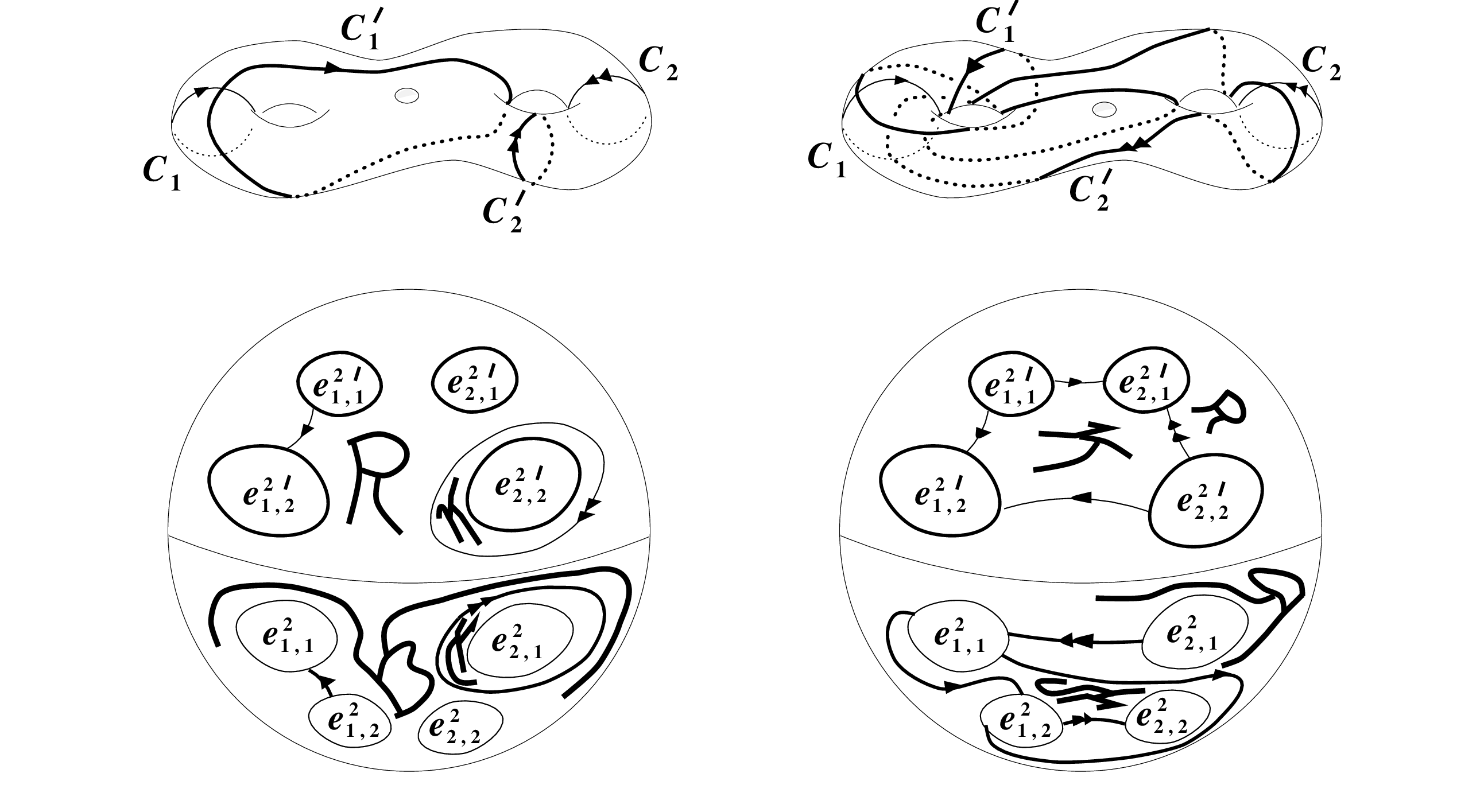,width=13cm,caption=}}
\end{figure}

\bigskip

\noindent
{\it (ii) 2-homotopy classes.}
Since $\mbox{\rm K3}$ is simply-connected, the set of map-classes from
$X^2$ to $\mbox{\rm K3}$ is enumerated by
$H^2(X^2;\pi_2(\mbox{\rm K3}))=H^2(M^3;\pi_2(\mbox{\rm K3}))$. On the
other hand, every map $f$ from $X^2$ to K3 extends to $M^3$ since the
composition $f\circ h$ from $\partial\,e^3$ to K3 is always
null-homotopic. Hence the set $[M^3,\mbox{\rm K3}]_2$ of $2$-homotopy
classes of maps from $M^3$ to K3 is also enumerated by
$H^2(M^3;\pi_2(\mbox{\rm K3}))$.

\bigskip

\noindent
{\it (iii) Map-classes.}
Given a $2$-homotopy class of maps from $M^3$ to $\mbox{\rm K3}$
represented by $f$ and labelled by $w$ in
$H^2(M^3;\pi_2(\mbox{\rm K3}))$. We shall first try to figure out
the subgroup $J_w$ in $H^3(M^3;\pi_3(\mbox{\rm K3}))$ and then
the subset $A_f$ in $H^3(M^3;\pi_3(\mbox{\rm K3}))/J_w$.

Up to homotopy, we may assume that the image of $f$ lies in
$\vee_{22}\, S^2$ with $\Sigma_g^{\circ}$ all mapped to the vertex
$\ast$ and that $f$ is smooth on the open subset of $M^3$ that is
mapped to the complement of $\ast$. Observe that there is a natural
surjection from $\pi_1(\mbox{\it Map}\,(X^2,\vee_{22}\,S^2), f|_{X^2})$
onto $\pi_1(\mbox{\it Map}\,(X^2,\mbox{\rm K3}),f|_{X^2})$ since every
loop in $\mbox{\it Map}\,(X^2,\mbox{\rm K3})$ at $f|_{X^2}$ is
homotopic to a loop in $\mbox{\it Map}\,(X^2,\vee_{22}\,S^2)$ relative
to the base point $f|_{X^2}$. Recall also that
$$
 H^3(M^3;\pi_3(\,\cdot\,))\;
 =\; \mbox{\it Hom}\,(H_3(M^3;{\Bbb Z})\,,\,\pi_3(\,\cdot\,))\;
 =\; \pi_3(\,\cdot\,)
$$
since $H_2(M^3;{\Bbb Z})$ is free. Consequently, $J_w$ in
$H^3(M^3;\pi_3(\mbox{\rm K3}))$ is the quotient of $J_w$ in
$H^3(M^3;\pi_3(\vee_{22}\,S^2))$ by $[h_0]$ and we only need
to study the latter.

Consider now a map $F:X^2\times S^1\rightarrow \vee_{22}\,S^2$ with
$F|_{X^2\times\{0\}}=f|_{X^2}$, which describes a loop in
$\mbox{\it Map}\,(X^2,\vee_{22}\,S^2)$ at the point $f|_{X^2}$.
As a homotopy from $f|_{X^2}$ to itself, $F$ then defines an obstruction
cohomology class $\delta^3(f,f;F)$ in $H^3(M^3;\pi_3(\vee_{22}\,S^2))$.

\bigskip

\noindent
{\bf Lemma 2.2.} {\it Let $W^2$ be the $2$-subcomplex in
 $X^2\times S^1$ realized by
 $$
  (\Sigma_g^{\circ}\times\{0\})\;\mbox{$\bigcup$}\;
         (\partial\Sigma_g^{\circ}\,
           \cup\, \cup_{i=1}^{g}\, C_i\,
            \cup\, \cup_{i=1}^{g}\, C_i^{\prime})\times S^1\,.
 $$
 Then $F$ can be homotoped so that $F$ sends $W^2$ to the vertex
 $\ast$ of $\vee_{22}\,S^2$, in addition to the requirement that
 $F|_{X^2\times \{0\}} = f|_{X^2}$. }

\bigskip

\noindent
Indeed $F$ can be homotoped further so that it is smooth on the open
subset of $X^2\times S^1$ that is mapped to the complement of $\ast$.
And we shall call such $F$ {\it nice}.

\bigskip

\noindent
{\it Proof.} The second homology $H_2(W^2;{\Bbb Z})$ of $W^2$ is
generated by the $2$-cycles: $\partial\Sigma_g^{\circ}\times S^1$
and $C_i\times S^1$, $C_i^{\prime}\times S^1$, $i=1,\cdots, g$. Since
each of $\partial\Sigma^{\circ}$ and $C_i$, $C_i^{\prime}$ bounds a
$2$-submanifold in $X^2$, each of the generating $2$-cycles bounds a
$3$-submanifold in $X^2\times S^1$ and hence is homologous to $0$ in
$H_2(X^2\times S^1;{\Bbb Z})$. Thus the induced map $(F|_{W^2})_{\ast}$
from $H_2(W^2;{\Bbb Z})$ to $H_2(\vee_{22}\,S^2;{\Bbb Z})$ is a zero
map. Since $H_1(W^2;{\Bbb Z})$ is torsion free and $\vee_{22}\,S^2$ is
simply-connected, the Hopf-Whitney classification theorem then implies
that $F|_{W^2}$ is homotopic to the constant map that takes $W^2$ to
$\ast$. By the homotopy extension property of polyhedral pairs, one
can extend this homotopy first to
$(\{e^2_i,e^{2\,\prime}_i\}_{i=1}^g\times\{0\})\times [0,1]$ via
$f|_{e^2_i}$ and $f|_{e^{2\,\prime}_i}$; and then to the rest
of $(X^2\times S^1)\times [0,1]$. This completes the proof.

\noindent \hspace{14cm} $\Box$

\bigskip

Fix a decomposition of $S^3$
$$
 S^3\;=\;e^3_-\,\cup\, (S^2\times[0,1])\, \cup\, e^3_+\,.
$$
Up to homotopy, there is a unique (continuous) map $\tau$ from $S^3$
to $M^3$ that satisfies: (1) $\tau$ is a diffeomorphism on the interior
of each of $e^3_-$, $e^3_+$ onto the interior of $e^3$, and
(2) $\tau|_{S^2\times[0,1]}=h|_{\partial\,e^3}$ via the projection from
$S^2\times [0,1]$ to $S^2$. The class $\delta^3(f,f;F)$
can then be realized as a map $G_{(f,f;F)}$ from $S^3$ to
$\vee_{22}\,S^2$, defined by
$$
 G_{(f,f;F)}\;=\;\left\{
         \begin{array}{ccl}
                f\circ\tau  &  \mbox{on} & e^3_-\cup e^3_+ \\[1ex]
                   F\circ (h\times \mbox{\it Id})
                            & \mbox{on} & S^2\times [0,1]
                  \end{array}\right.\;,
$$
where $\mbox{\it Id}$ is the identity map of the interval $[0,1]$.

Up to homotopy, one may assume that $F$ is nice. By Sard's theorem
there exist common regular values of $f$ and $F$:
$y_i$, $i=1,\cdots, 22$, one on each oriented $S^2$ in the bouquet.
Since the preimage $F^{-1}(y_i)$ is now an oriented framed 1-manifold,
with possibly several components, contained in the interior of the
submanifold
$$
 X^2\times S^1 - W^2\;
 =\;(X^2-\partial\Sigma_g^{\circ}-
    \cup_{i=1}^{g}C_i-\cup_{i=1}^{g}C_i^{\prime})\times S^1\;
     -\; \Sigma_g^{\circ}\times\{0\}
$$
in $X^2\times S^1$, the preimage $G_{(f,f;F)}^{-1}(y_i)$ is then a
well-defined oriented framed link $L_i$ in $S^3$. The linking matrix
$(\mbox{\it lk}\,(L_i,L_j))_{ij}$ indicates then what class
$G_{(f,f;F)}$ represents in
$H^3(M^3;\pi_3(\vee_{22}\,S^2))=\pi_3(\vee_{22}\,S^2)$.

Let us now take a closer look at this collection of links in $S^3$.
Recall the map $\rho$ on $S^2$. It extends naturally to
$S^2\times [0,1]$ via the product structure. We shall denote this new
map still by $\rho$. Then, due to niceness of $F$ and the construction,
the framed 1-manifolds $L_i\cap(S^2\times[0,1])$, without orientation,
are invariant under $\rho$ with their boundary lying in the interior of
the $2$-discs
$$
 \left(\{\,e^2_{i,1}\,,\,e^2_{i,2}\,\}_{i=1}^g\times\{0,1\}\right)\,
  \mbox{$\bigcup$}\, \left(\{\,e^{2\,\prime}_{i,1}\,,\,
         e^{2\,\prime}_{i,2}\,\}_{i=1}^g\times\{0,1\}\right)\,.
$$
This renders a component knot $K$ of the links be of only two types:
\begin{quote}
 \parbox[t]{5em}{Type (a):} \
  \parbox[t]{12cm}{$K$ lies in the interior of $\Omega\times [0,1]$,
     where $\Omega$ is some $\Omega_j$ or $\Omega_j^{\prime}$. }

 \parbox[t]{5em}{Type (b):} \
  \parbox[t]{12cm}{$K$ lies in the interior of the handlebody of genus
     $4g-1$,

     $$
      e^3_-\,\cup\,
       \left(\{\,e^2_{i,1}\,,\, e^2_{i,2}\,,\,
         e^{2\,\prime}_{i,1}\,,\, e^{2\,\prime}_{i,2}\,\}_{i=1}^g
               \times [0,1]\right)\, \cup \, e^3_+\,.
     $$ }
\end{quote}
Consequently, one can {\it split} the links
$$
 L_i\;\longrightarrow\;L_i^-\,\cup\, L_i^0\,\cup\, L_i^+\,
$$
through a collection of disjoint segments in $S^2\times\{0,1\}$ as
indicated in {\sc Figure 2.2}. Each of these segments as viewed from
$\Sigma_g^{\circ}$ is a dual loop to either some $C_i$ in
$\{C_1,\cdots, C_g\}$ or some $C_i^{\prime}$ in
$\{C_1^{\prime}, \cdots, C_g^{\prime}\}$. After being smoothed if
necessary, the oriented framed links satisfy
$$
 \mbox{\it lk}\,(L_i,L_j) \;
   =\;\mbox{\it lk}\,(\,L_i^- \cup L_i^0 \cup L_i^+\,,\,
         L_j^- \cup L_j^0 \cup L_j^+\,)\,.
$$

\begin{figure}[htbp]
\setcaption{{\sc Figure 2.2.} The splitting of $L_i$ into
    $L_i^-\cup L_i^0 \cup L_i ^+$. Up to homotopy, each component knot
    $K$ of Type (b) is now contained in a solid torus $\Theta_i$ or
    $\Theta_i^{\prime}$. Those component knots of Type (a) are
    indifferent of such splittings and hence are omitted for the clarity
    of the picture. Notice that the $\rho$ symmetry is maintained after
    the splitting and the collection of links in $e^3_+$ and those in
    $e^3_-$ are now the mirror-image of each other.  }
\centerline{\psfig{figure=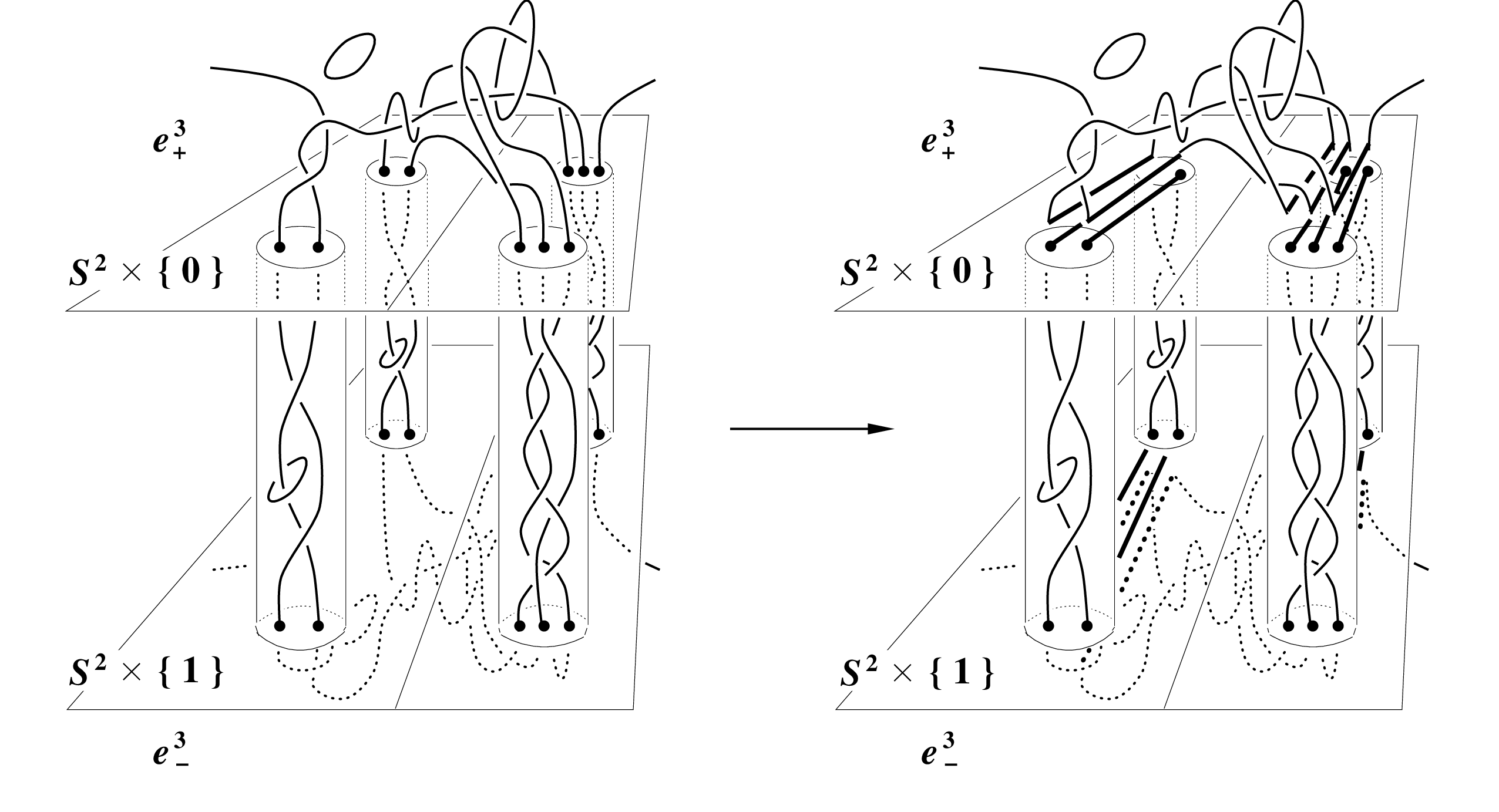,width=13cm,caption=}}
\end{figure}

After the splitting, each component knot $K$ of Type (b) is either
contained in $e^3_-$, $e^3_+$, or a solid torus $\Theta_i$,
$\Theta_i^{\prime}$ (cf.\ {\sc Figure 2.3}). The map $\rho$ can be
extended to each $\Theta_i$ and $\Theta_i^{\prime}$, on which it is an
orientation-reversing homeomorphism. By construction, $\rho$ leaves the
links in a $\Theta_i$ or $\Theta_i^{\prime}$ invariant with the framing
preserved while the orientation reversed. The two collections of
oriented framed links $\{L_1^-,\cdots,L_{22}^-\}$ and
$\{L_1^+,\cdots,L_{22}^+\}$ are contained in two separate $3$-balls
$B^3_-$ and $B^3_+$ that are disjoint from any of the $L_j^0$.
The reflection between $e^3_-$ and $e^3_+$ with respect to the equator
$S^2\times\{\frac{1}{2}\}$ in the decomposition of $S^3$ induces an
orientation-reversing homeomorphism between $B^3_-$ and $B^3_+$ that
sends $^-$-links and $^+$-links to each other with the framing preserved
while the orientation reversed. Together with the fact that the linking
number of a pair of links remains the same if the orientation of both
links are reversed and it differs exactly by a sign from that of their
mirror-image pair, one concludes that
$$
 \mbox{\it lk}\,(L_i,L_j)\;
   =\; \mbox{\it lk}\,(L_i^-,L_j^-)\,
           +\,\mbox{\it lk}\,(L_i^0,L_j^0)\,
                    +\,\mbox{\it lk}\,(L_i^+,L_j^+) \;
   =\; \mbox{\it lk}\,(L_i^0,L_j^0)\,.
$$
Note that this is expected from the fact that $J_w$ depends only
on $w$, not on any particular $f$ that represents $w$.

Let $K_1$, $K_2$ be component knots of some $L_{i_0}^0$, $L_{j_0}^0$
and $\Omega$ (resp.\ $\Theta$) be some $\Omega_i$ or
$\Omega^{\prime}_i$ (resp.\ $\Theta_i$ or $\Theta^{\prime}_i$). {}From
the topological relations among regions $\Omega\times[0,1]$ and
$\Theta$ in $S^3$, there are only two situations that $K_1$, $K_2$ may
contribute to $\mbox{\it lk}\,(L_{i_0}^0,L_{j_0}^0)$:
\begin{quote}
  Case (1) \hspace{1em}
  Both $K_1$ and $K_2$ lie in a same $\Omega\times[0,1]$ or $\Theta$,

  Case (2) \hspace{1em}
  Say, $K_1$ lies in an $\Omega\times[0,1]$ and $K_2$ lies in a
  $\Theta$.
\end{quote}
In Case (1), the total contribution of such $(K_1,K_2)$ to
$\mbox{\it lk}\,(L_{i_0}^0,L_{j_0}^0)$ vanishes due to the invariance
of $(L_{i_0}^0,L_{j_0}^0)$ under $\rho$ up to the overall reversing of
orientations. Consequently, the only contributions to the total
linking number are from $(K_1,K_2)$ in Case (2). In general this can be
still complicated to analyze. However, there are simpler $M^3$, for
which the task is straightforward.

Recall that an orientable $3$-manifold $M^3$ is called {\it prime} if
any direct sum decomposition $M^3=M_1^3\,\sharp\,M^3_2$ implies that
either $M^3_1$ or $M^3_2$ is an $S^3$; and is called {\it irreducible}
if every embedded $S^2$ bounds a $3$-ball ([He]). The only prime,
non-irreducible orientable $3$-manifold is $S^2\times S^1$.
A $3$-manifold is {\it reducible} if it is not prime.

Suppose that $M^3$ is irreducible and that the Heegaard splitting in
the discussion has the minimal genus. Consider $K_1$, $K_2$ in
Case (2).

\bigskip

\noindent
{\bf Lemma 2.3.} {\it Let $M^3$ be an irreducible closed orientable
 $3$-manifold with a Heegaard splitting of minimal genus
 $$
  M^3\;=\; H_g\,\cup_{\Sigma_g}\,H_g^{\prime}\,.
 $$
 Then every loop $\gamma$ in a component $\Xi$ of the complement of
 $\{C_i, C_i^{\prime}\}_{i=1}^g$ in $\Sigma_g$ is null-homotopic in
 $\Xi$. In other words, $\Xi$ is a disc. }

\bigskip

\noindent
{\it Proof.} We only need to consider the case $\gamma$ is a simple
loop in $\Xi$. Since $\gamma$ is contained in the complement of all
$C_i$ and $C_i^{\prime}$, it is null-homotopic in both $H_g$ and
$H_g^{\prime}$. By Dehn's Lemma [He], $\gamma$ bounds an embedded
$2$-disc $\Delta^2$ in $H_g$ and an embedded $2$-disc
$\Delta^{2\,\prime}$ in $H_g^{\prime}$. Together they give an embedded
$2$-sphere $S^2$ in $M^3$. Since $M^3$ is irreducible, this $S^2$ bounds
a $3$-ball $B^3$. The intersection $B^3\cap\Sigma_g$ must then be a
$2$-disc in $\Xi$ with boundary $\gamma$ since otherwise the genus of
the Heegaard splitting can be reduced, contradicting the minimal genus
assumption. Therefore, $\gamma$ is null-homotopic in $\Xi$. This
concludes the lemma.

\noindent \hspace{14cm} $\Box$

\bigskip

\noindent
Consequently, if $\Omega$ does not contain
$h^{-1}(\partial\Sigma_g^{\circ})$ in its boundary, then $K_1$ does not
link around $K_2$ at all and hence $\mbox{\it lk}\,(K_1,K_2)=0$. If, on
the other hand, $\Omega$ contains $h^{-1}(\partial\Sigma_g^{\circ})$ in
its boundary, then, under homotopy, one may assume that $K_1$ lies in a
small neighborhood of $h^{-1}(\partial\Sigma_g^{\circ})\times[0,1]$ and
hence again does not link around $K_2$ either. This shows that the
linking matrix associated to $G_{(f,f;F)}$ for any nice $F$ is in fact
the $22\times 22$ zero matrix; and $G_{(f,f;F)}$ represents the zero
class in $\pi_3(\vee_{22}\,S^2)$.

\bigskip

\noindent
{\bf Corollary 2.4.}
{\it If $M^3$ is irreducible, then $J_w =0$ for every $w$ in
 $H^2(M^3;\pi_2(\mbox{\rm K3}))$. }

\bigskip

For $M^3=M^3_1\,\sharp\,M^3_2$, there is a natural binary
operation: ($\,\cdot\,$ $=$ either K3 or $\vee_{22}\,S^2$.)
$$
\begin{array}{cccccl}
 + & : & H^3(M^3_1;\pi_3(\,\cdot\,))
           \oplus H^3(M^3_2;\pi_3(\,\cdot\,))
   & \longrightarrow  & H^3(M^3;\pi_3(\,\cdot\,)) \\[1ex]
   & & \left(\,([M^3_1]\mapsto\alpha_1)\,,\,
                  ([M^3_2]\mapsto\alpha_2)\,\right)
        & \longmapsto  & (\,[M^3]\mapsto(\alpha_1+\alpha_2)\,) &,
\end{array}
$$
which commutes with the identification of each of the
$H^3(\cdot\cdot\,;\pi_3(\,\cdot\,))$ with $\pi_3(\,\cdot\,)$.
Under this operation, one has

\bigskip

\noindent
{\bf Lemma 2.5.}
{\it Suppose that $M^3$ can be decomposed into a connected sum
 $M^3_1\,\sharp\,M^3_2$. Let $w=w_1+w_2$ following the decomposition
 $$
  H^2(M^3;\pi_2(\mbox{\rm K3}))\;
    =\;H^2(M^3_1;\pi_2(\mbox{\rm K3}))\,
          \oplus\, H^2(M^3_2;\pi_2(\mbox{\rm K3}))\,.
 $$
 Then
 $$
   J_w\;=\;J_{w_1}\,+\, J_{w_2}\,.
 $$
} 

\bigskip

\noindent
{\it Proof.} Let $M^3_i=X^2_i\cup_{h_i}e^3_i$, $i=1,\,2$,
be a CW-complex structure of $M^3_i$ induced from a Heegaard splitting.
Let $M^3=X^3\cup_h e^3$ be a compatible CW-complex structure for $M^3$.
Then the relation between Heegaard splitting and connected sum, together
with the discussion in this section, implies that the reduced oriented
framed link $L_i^0$ in $S^3$ indeed separates into two parts
$L_{i,1}^0$ and $L_{i,2}^0$: the former corresponds to the
$M^3_1$-component and the latter to the $M^3_2$-component; and each is
contained in a $3$-ball (cf.\ {\sc Figure 1.2}). Hence,
$$
 \mbox{\it lk}\,(L_i^0,L_j^0)\;
   =\;\mbox{\it lk}\,(L_{i,1}^0,L_{j,1}^0)\,
          +\,\mbox{\it lk}\,(L_{i,2}^0,L_{j,2}^0)\,.
$$
The lemma thus follows.

\noindent \hspace{14cm} $\Box$

\bigskip

\noindent
Consequently, $J_w$ vanishes for every $w$ in
$H^2(M^3;\pi_2(\mbox{\rm K3}))$ if $M^3$ is the connected sum of a
finite collection of irreducible orientable $3$-manifolds.

If one decomposes a general $M^3$ into
$$
 M^3\;=\; M^3_0\,\sharp\, c\,(S^2\times S^1)\,,
$$
where the prime decomposition of $M^3_0$ contains no $S^2\times S^1$.
Recall from [Mi2] that both $M^3_0$ and $c$ are determined by $M^3$.
Decompose $H^2(M^3;\pi_2(\mbox{\rm K3}))$ correspondingly as
$H^2(M^3_0;\pi_2(\mbox{\rm K3}))
 \bigoplus \oplus_c\, H^2(S^2\times S^1;\pi_2(\mbox{\rm K3}))$
and $w$ as $w_0+ w_1+\cdots + w_c$. Then
$$
 J_w\; =\; J_{w_1} + \cdots + J_{w_c}\,.
$$
This somehow distinguishes the $3$-manifold $S^2\times S^1$ in the
current problem.

\bigskip

\noindent
{\bf Lemma 2.6.} {\it Let $M^3=S^2\times S^1$ and
 $w=(k_1,k_2,\cdots, k_{22})$ be in
 $H^2(S^2\times S^1;\pi_2(\mbox{\rm K3}))=\oplus_{22}\,{\Bbb Z}$.
 Then the abelian subgroup $J_w$ in $\pi_3(\vee_{22}\,S^2)$ is
 generated by
 $$
  \mbox{\boldmath $u$}_{\,i}\;
   =\; k_i [i]\,+\,\sum_{j<i}\,k_j [j,i] \,+\,\sum_{j>i}\,k_j [i,j]
      \hspace{1em} \sim  \hspace{1em}
       \left( \begin{array}{cclcc}
                        &         & k_1      &        & \\
                        &         & \vdots   &        & \\
                  k_1   &\cdots   & k_i      & \cdots & k_{22} \\
                        &         & \vdots   &        & \\
                        &         & k_{22}   &        & \\
              \end{array} \right)\;,
 $$
 for $i=1,\cdots, 22$, where, for each $i$, the entries in the
 corresponding linking matrix are all zero except those in
 the $i$-th column or the $i$-th row. In particular, $J_w$ is trivial
 for $w=0$ and is of rank $22$ for $w\neq 0$.  }

\bigskip

\noindent
{\it Proof.} A Heegaard splitting of $S^2\times S^1$ and its associate
decomposition of $S^2$ are indicated in {\sc Figure 2.3}. Following the
notations introduced in Sec.\ 1, note that
$H_2(X^2;{\Bbb Z})\cong {\Bbb Z}$ is generated by
$e^2_1-e^{2\,\prime}_1$. One may assume then that $f^{-1}(y_j)$ when
restricted to $X^2$ consists of $k_j$ distinct points in the interior
of $e^2_1$ and none in the interior of $e^{2\,\prime}_1$. Subject to
this, by the Pontryagin-Thom construction ([Mi3]), any oriented framed
link as indicated in {\sc Figure 2.3} is realizable by some map $f$
from $S^2\times S^1$ to $\vee_{22}\,S^2$ in the given $2$-homotopy
class. The linking matrices and hence their corresponding
elements in $\pi_3(\vee_{22}\,S^2)$ are exactly
$\mbox{\boldmath $u$}_i$. Since only component knots in Case (2)
contribute to the linking matrix, the combinations of integral copies
of $K_{(i)}$ as shown, which is part of the preimage $f^{-1}(y_i)$,
exhaust all the possible linking matrices associated to maps in the
$2$-homotopy class $w$. This shows that
$\mbox{\boldmath $u$}_{\,1}$, $\cdots$, $\mbox{\boldmath $u$}_{\,22}$
form a generating set for $J_w$. It is straightforward to check that
they are linearly independent unless $w=0$. This concludes the lemma.

\begin{figure}[htbp]
\setcaption{{\sc Figure 2.3.} A Heegaard splitting of $S^2\times S^1$
  by a torus and the corresponding decomposition of $S^2$ are indicated.
  The collection of links in the solid torus $\Theta_1$ depends on the
  $2$-homotopy class $w$. The knot $K_{(i)}$ in the preimage
  $f^{-1}(y_i)$ gives $\mbox{\boldmath $u$}_{\,i}$. Together they
  generate $J_w$.}
\centerline{\psfig{figure=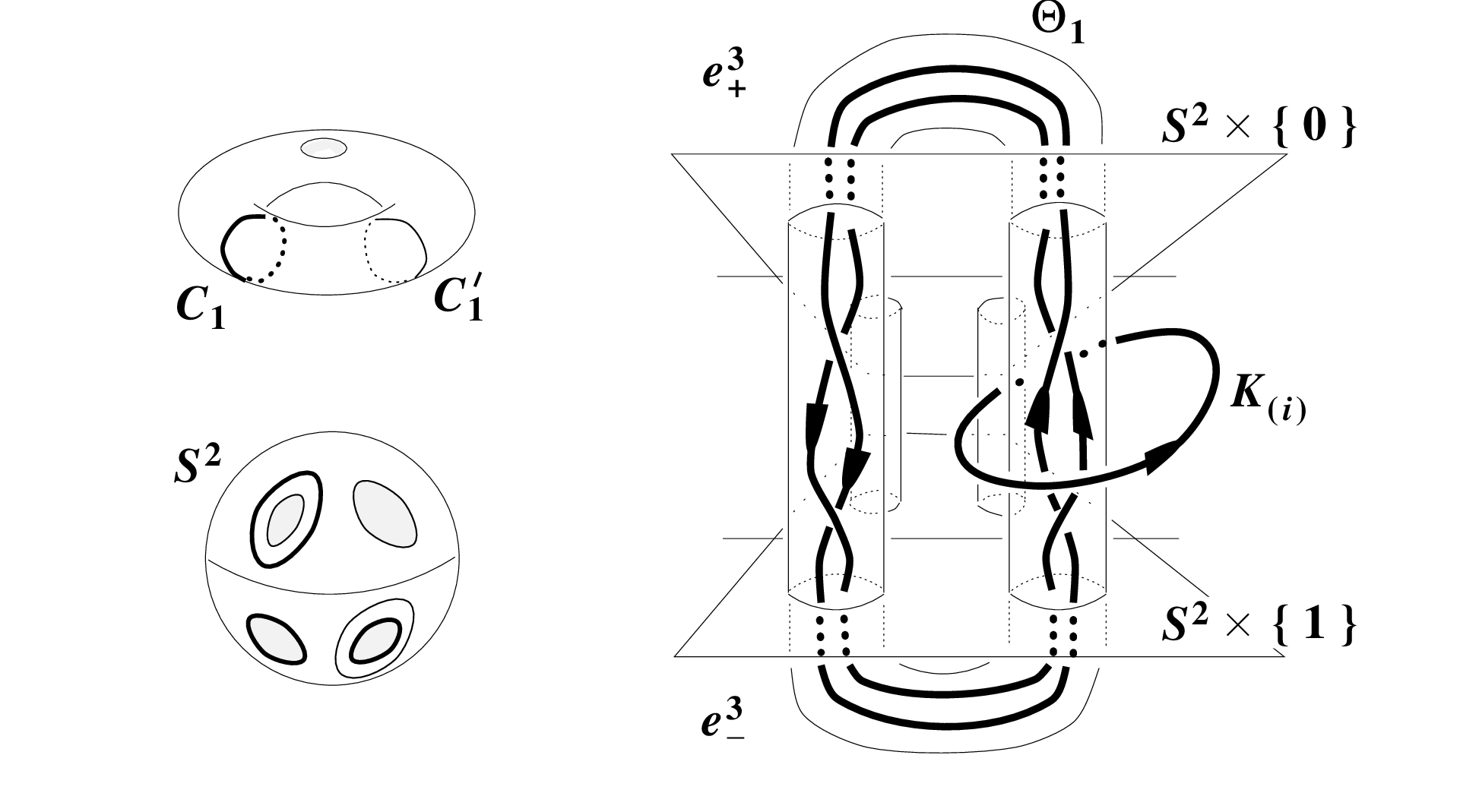,width=13cm,caption=}}
\end{figure}

\noindent \hspace{14cm} $\Box$

\bigskip

This concludes our discussion on $J_w$-part. As for $A_f$-part, notice
that

\bigskip

\noindent
{\bf Lemma 2.7.} {\it Fix an $f_0$ in $w$. Then every element in
 $H^3(M^3;\pi_3(\mbox{\rm K3}))$ can be represented by the obstruction
 class $\delta^3(f_0,f;\,\cdot\,)$ for some $f$ in $w$. }

\bigskip

\noindent
{\it Proof.} Consider a trivial connected sum $M^3=M^3\,\sharp\,S^3$
whose involved $S^2$ is mapped to a point $\ast$ in K3 after a homotopy
of $f_0$, and the $3$-ball $B^3$ it bounds is disjoint from $X^2$. Let
$g$ be a map from $(B^3,S^2)$ to $(\mbox{\rm K3},\ast)$. Then
the $f$ defined on $M^3$ by amalgamating $f_0$ and $g$, following the
trivial connected sum $M^3\,\sharp\,S^3$, is $2$-homotopic to $f_0$
since $f_0|_{X^2}$ and $f|_{X^2}$ are identical. Now that $g$ can be
any class in $\pi_3(\mbox{\rm K3})$, one concludes the lemma.

\noindent \hspace{14cm} $\Box$

\bigskip

\noindent
{\it Remark 2.8.} Notice that, up to homotopy, every pair of
 $2$-homotopic maps $f_1$, $f_2$ from $M^3$ to K3 are related as in
 the above proof.

\bigskip

Consequently, one has
$$
 [M^3,\mbox{\rm K3}]\;
  =\; \bigcup_{[f_0]_2\in H^2(M^3;\pi_2(\rm K3))} A_{f_0} \;\;
  \simeq\; \bigcup_{w\in (\oplus_{22}\,{\subscriptsizeBbb Z})^{b_2}
   \oplus
  \oplus_{22}\mbox{\scriptsize\it Tor}\,H_1(M^3;{\subscriptsizeBbb Z})}
      \left.\mbox{\raisebox{1ex}{$\oplus_{253}\,{\Bbb Z}$}} \right/
            \mbox{\raisebox{-1ex}{$[h_0]+J_w$}} \;\;,
$$
where $b_2$ is the second Betti number of $M^3$ and
$\oplus_{253}\,{\Bbb Z}/([h_0]+J_w)$ should be regarded only as a
lattice (i.e.\ an integral affine space).

\bigskip

\noindent
{\bf Corollary 2.9.} {\it If the prime decomposition of $M^3$ does
 not contain any $S^2\times S^1$ as a component (- note that this holds
 if and only if $M^3$ contains no non-separating embedded $S^2$ -), then
 \begin{eqnarray*}
  \lefteqn{ [M^3,\mbox{\rm K3}]\;
               =\; H^2(M^3;\pi_2(\mbox{\rm K3}))
                      \times H^3(M^3;\pi_3(\mbox{\rm K3})) } \\
   & & \simeq\;
        \left( (\oplus_{22}\,{\Bbb Z})^{b_2}\,\mbox{$\bigoplus$}\,
         \oplus_{22}\,\mbox{\it Tor}\,H_1(M^3;{\Bbb Z})\right)\,
          \times\, \oplus_{252}\,{\Bbb Z}\,.
 \end{eqnarray*}
 }

\bigskip

All together this concludes the enumeration of topological membrane
solitons, represented by classes in $[M^3,N]$.

\bigskip

\noindent
{\it Remark 2.10.} For
 $N={\Bbb R}^{3+1}\times [0,1] \times {\Bbb T}^2\times \mbox{\rm K3}$
 with complex K\"{a}hler ${\Bbb T}^2\times \mbox{\rm K3}$, two
 immediate problems follow: (1) {\it How can one characterize those
 topological membrane solitons that contain physical, say BPS-,
 solitons?} (2) {\it What is the moduli space of these physical
 solitons?} Both remain unsolved at the moment.

\bigskip

\section{Patching of membrane solitons and topological charges.}

In this section, we shall focus on the specific string-admissible
target-space
$$
 N\; =\; {\Bbb R}^{3+1}\times [0,1] \times {\Bbb T}^2
          \times \mbox{\rm K3}\,,
$$
which should play some role in the development of M-theory. The
total dimension $11$ has the origin from supergravity; and
coincidentally the dimension $5$ of the contractible part
${\Bbb R}^{3+1}\times [0,1]$ is the minimal dimension for a real vector
space to allow every compact orientable $3$-manifold to be embedded
therein. Following from this, every class in $[M^3,N]$ can be
represented by an embedding.

\bigskip

\begin{flushleft}
{\bf Patching of solitons and additivity of charges.}
\end{flushleft}
Let us consider the patchings first.
Let $(M^3_1,p_1)$ and $(M^3_2,p_2)$ be two oriented closed
$3$-manifolds-with-base-point and let $f_i$ be a map from $M^3_i$ to
$N$ that represents a membrane soliton. If some $M^3_i$ has non-empty
boundary, then we require that $p_i$ lies in the interior of $M^3_i$.
Let $\gamma$ be a path in $N$ connecting $f_1(p_1)$ and $f_2(p_2)$.
One can then define a {\it patching of membrane solitons} in $N$ by
"{\it puffing}" the following joining of maps:
$$
 f_1\cup\gamma\cup f_2 \;:\;
    M^3_1\,\cup_{p_1\sim 0}\,[0,1]\,\cup_{1\sim p_2}\,M^3_2 \;
  \longrightarrow\; N\,.
$$
The result is an amalgamation $f_1\,\sharp_{\gamma}\,f_2$ from
the connected sum $M^3_1\,\sharp\,M^3_2$ into $N$. ({\sc Figure 3.1}.)

\begin{figure}[htbp]
\setcaption{{\sc Figure 3.1.} A map $f_1\,\sharp_{\gamma}\,f_2$ from
    $M^3_1\,\sharp\,M^3_2$ to $N$ obtained by "puffing" along $\gamma$.
    This defines a patching of solitons.  }
\centerline{\psfig{figure=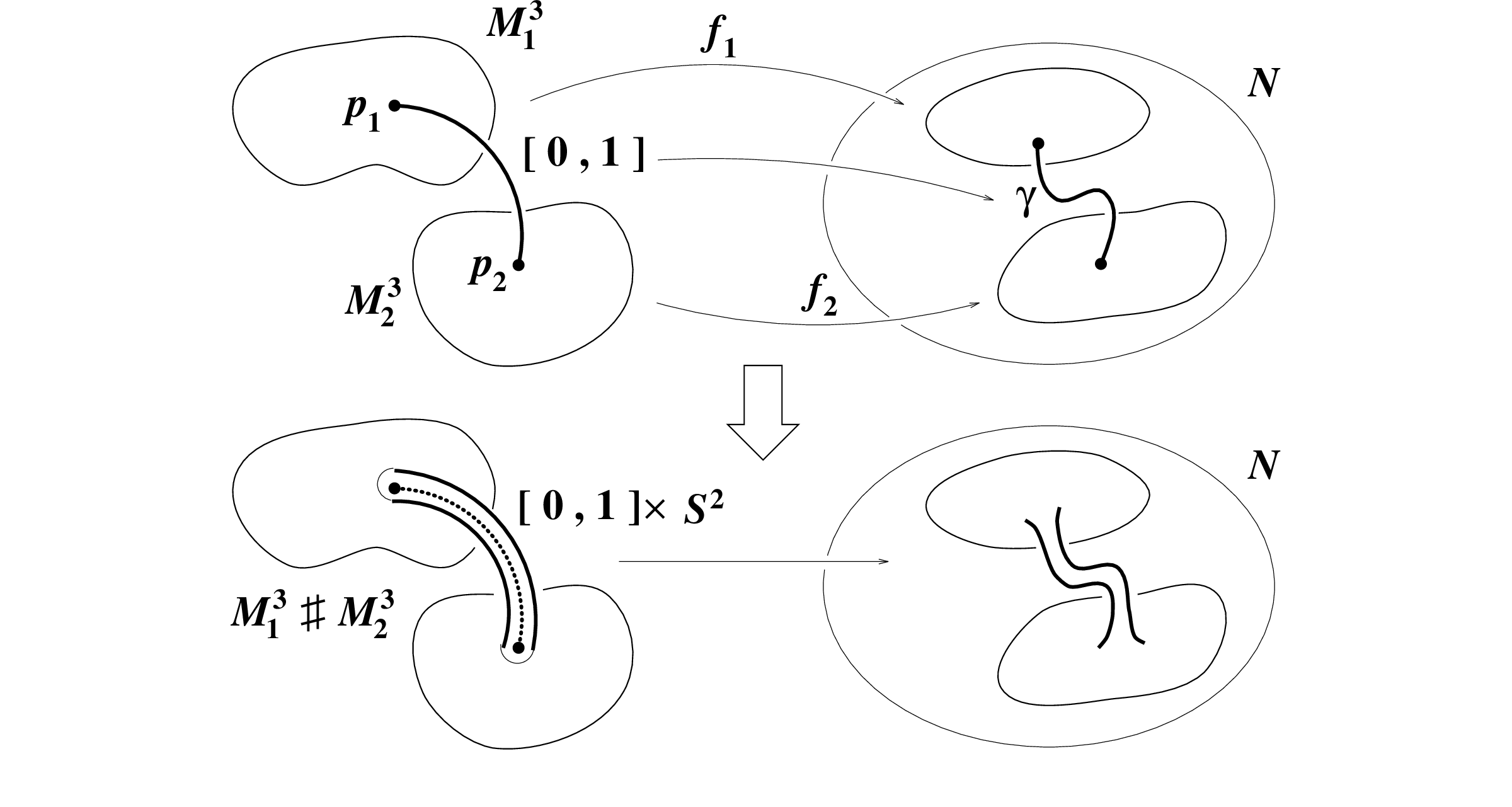,width=13cm,caption=}}
\end{figure}

In general, one would expect that different choices of $p_1$, $p_2$,
and $\gamma$ may lead to non-homotopic maps from $M^3_1\,\sharp\,M^3_2$
to $N$. It turns out that this is not the case for our $N$.
Let us look at the projection into the components ${\Bbb T}^2$ and K3.
Since $M^3_1$, $M^3_2$ are path-connected and
$\pi_1(M^3_1\,\sharp\,M^3_2)$ is the free product
$\pi_1(M^3_1)\ast\pi_1(M^3_2)$, the induced map
$(f_1\,\sharp_{\gamma}\,f_2)_{\ast}$ is independent of the choices of
$p_1$, $p_2$, and $\gamma$. Hence, for the ${\Bbb T}^2$-component,
$f_1\,\sharp_{\gamma}\,f_2$ gives a well-defined homotopy class
$[f_1\,\sharp\,f_2]$ in $[M^3_1\,\sharp\,M^3_2,{\Bbb T}^2]$ since
the latter is $H^1(M^3_1\,\sharp\,M^3_2;\pi_1({\Bbb T}^2))$, which is
$\mbox{\it Hom}\,(H_1(M^3_1\,\sharp\,M^3_2;{\Bbb Z})\,,\,
  H_1({\Bbb T}^2;{\Bbb Z}))$. For the K3-component, since K3 is
simply-connected, the same conclusion holds. Altogether, one has a
well-defined class $[f_1\,\sharp\,f_2]$ in $[M^3_1\,\sharp\,M^3_2,N]$.
In other words, for our $N$, the patching thus defined is indeed an
operation at the topological level.

Let us now consider the charges. For $M^3$ closed, a map class $[f]$ in
$[M^3, N]$ defines a class $\langle f\rangle$ in $H_3(N;{\Bbb Z})$.
Closed $3$-forms on $N$, and hence classes in the de Rham cohomology
$H^3_{DR}(N)$, can then be eveluated on $\langle f\rangle$ and define
various {\it topological charges} for the soliton $f$. By construction,
$$
 \langle\,f_1\,\sharp\,f_2\,\rangle\;
  =\; \langle f_1\rangle\, +\, \langle f_2\rangle\,.
$$
And hence topological charges add under patchings.

For solitons with non-empty boundary, topological charges can also be
defined for those whose boundary lies entirely in $\partial N$.
Closed $3$-forms on $N$ that vanish at $\partial N$, and hence classes
in the restricted relative de Rham cohomology
$H_{DR}^{(r)\,3}(N,\partial N)$, can be evaluated on $\langle f\rangle$
for such $f$ and give the topological charges. Observe that
$$
 \left(H_{DR}^1([0,1],\{0,1\})
    \otimes H_{DR}^2({\Bbb T}^2\times\mbox{\rm K3})\right)
    \mbox{$\bigoplus$}\, H_4({\Bbb T}^2\times\mbox{\rm K3};{\Bbb R})\;
  \subset\; H_{DR}^{(r)\,3}(N,\partial N)\,;
$$
so there are actually a plenty source of charges for boundary-confined
solitons in $(N,\partial N)$. The additivity property of charges under
patchings still holds.

As a comparison, the patching of membrane solitons defined here
generalizes the patching that occurs in gauge theory. In the latter
case, homotopy groups and, hence, amalgamations of maps from
connected-sums of spheres are involved ([Co]). The fact that topological
charges add under patchings indicate that such generalization is natural
for $N$.

Since topological charges depend only on $\langle f\rangle$, the
information is contained entirely in the {\it tautological map}
either from $[(M^3,\partial M^3),(N,\partial N)]$ to
$H_3(N,\partial N;{\Bbb Z})$ if $\partial M^3$ is non-empty, or from
$[M^3,N]$ to $H_3(N;{\Bbb Z})$ if $M^3$ is closed, that sends $[f]$
to $\langle f\rangle$. We shall now study some details of this
correspondence.

\bigskip

\begin{flushleft}
{\bf The tautological map: (1) When $\partial M^3$ is nonempty.}
\end{flushleft}
Let
$\partial N
 = {\Bbb R}^{3+1}\times\{0,1\}\times{\Bbb T}^2\times\mbox{\rm K3}
 = \partial_0 N \cup \partial_1 N$.
{}From the contractibility of ${\Bbb R}^{3+1}$ and $[0,1]$, and the
Poincar\'{e} and the Poincar\'{e}-Lefschetz dualities, one has
\begin{eqnarray*}
 \lefteqn{ H_3(N,\partial N ;{\Bbb Z})\;
  =\; H^4([0,1]\times{\Bbb T}^2\times\mbox{\rm K3} ;{\Bbb Z})\;
  =\; H_2({\Bbb T}^2\times\mbox{\rm K3} ;{\Bbb Z}) }\\
  & &  =\; H_2({\Bbb T}^2 ;{\Bbb Z})
                   \oplus H_2(\mbox{\rm K3} ;{\Bbb Z})\;
       =\; H_2(\partial_0 N;{\Bbb Z})\;
       =\; H_2(\partial_1 N;{\Bbb Z})\,.
\end{eqnarray*}
As a result, $H_3(N,\partial N ;{\Bbb Z})$ is generated by the
embedded $3$-manifolds-with-boundary $[0,1]\times{\Bbb T}^2$ and
$[0,1]\times S^2_j$, $j=1, \cdots, 22$.

Let $\partial M = \Sigma^2_{(1)} \cup \cdots \cup \Sigma^2_{(r)}$
and $f:(M^3,\partial M^3)\rightarrow (N,\partial N)$. If
$f(\partial M)$ is contained completely in, say, $\partial_0 N$,
then $f$ can be homotoped into $\partial_0 N$, thus
$\langle f\rangle=0$ in $H_3(N,\partial N;{\Bbb Z})$. In general,
up to a relabelling, suppose that $f$ maps the first $r_0$ components
of $\partial M$ into $\partial_0 N$ and the rest into $\partial_1 N$.
Then, under the above isomorphism,
\begin{eqnarray*}
 \langle f \rangle \;\;\mbox{in $H_3(N,\partial N ;{\Bbb Z})$}
  & = & \langle f|_{\Sigma^2_{(r_0+1)}} \rangle + \cdots
             + \langle f|_{\Sigma^2_{(r)}} \rangle\;\;
                     \mbox{in $H_2(\partial_1 N ;{\Bbb Z})$} \\[1ex]
  & = & - \langle f|_{\Sigma^2_{(1)}} \rangle - \cdots
             - \langle f|_{\Sigma^2_{(r_0)}} \rangle\;\;
                     \mbox{in $H_2(\partial_0 N ;{\Bbb Z})$}
\end{eqnarray*}
since such $f$ differs from one that takes $\partial M$ all into
$\partial_0 N$ by a dragging of
$\Sigma^2_{(r_0+1)}$, $\cdots$, $\Sigma^2_{(r)}$ from $\partial_0 N$
to $\partial_1 N$. Therefore, if let
$\mbox{\it pr}_1$, $\mbox{\it pr}_2$ be the projection maps from $N$ to
${\Bbb T}^2$ and K3 respectively, then
\begin{eqnarray*}
\lefteqn{ \langle f \rangle \;
   =\; \left(
    \langle (\mbox{\it pr}_1\circ f)|_{\Sigma^2_{(r_0+1)}}\rangle
  + \cdots +
   \langle(\mbox{\it pr}_1\circ f)|_{\Sigma^2_{(r)}}\rangle\right)}\\
 & & \hspace{6em}  \oplus \,
   \left(\langle(\mbox{\it pr}_2\circ f)|_{\Sigma^2_{(r_0+1)}}\rangle
  + \cdots +
   \langle(\mbox{\it pr}_2\circ f)|_{\Sigma^2_{(r)}}\rangle\right)\,.
\end{eqnarray*}
{}From this expression, one can obtain $\langle f\rangle$ from $[f]$
as follows.

Let $[{\Bbb T}^2]$ be the fundamental class of ${\Bbb T}^2$ in
$H_2({\Bbb T}^2;{\Bbb Z})$. Then the first summand in the above
expression is the summation of
$\mbox{\rm deg}\,(\mbox{\it pr}_1\circ f)\,\cdot\,[{\Bbb T}^2]$
from the involved component $\Sigma^2$ of $\partial M^3$ to
${\Bbb T}^2$. This can be computed, through Lemma 3.1 below, from
$i_{\ast}H_1(\Sigma^2;{\Bbb Z})$, the image of $H_1(\Sigma^2;{\Bbb Z})$
in $H_1(M^3;{\Bbb Z})$ under the inclusion map $i$ from $\partial M^3$
to $M^3$, and any $1$-cocycle of $M^3$ with coefficient in
$\pi_1({\Bbb T}^2)$ that represents $[\mbox{\it pr}_1\circ f]$.
Similarly, the second summand can be computed from $i_{\ast}(\Sigma)$
in $H_2(M^3;{\Bbb Z})$ and any $2$-cocycle of $M^3$ with coefficient in
$\pi_2(\mbox{\rm K3})$ that represents $[\mbox{\it pr}_2\circ f]$.
Explicit formula depends on the detail of the inclusion map $i$.

\bigskip

\noindent
{\bf Lemma 3.1.} {\it Let $f$ be a map from a closed oriented surface
 $\Sigma$ to an oriented torus ${\Bbb T}^2$. Let
 $(a_1, b_1, \cdots, a_g, b_g)$ be a canonical basis for
 $H_1(\Sigma;{\Bbb Z})$. Suppose that, with respect to a basis for
 $H_1({\Bbb T}^2;{\Bbb Z})$, $f_{\ast}(a_i)=(k_i,l_i)$ and
 $f_{\ast}(b_i)=(m_i,n_i)$, then the degree of $f$ is determined by
 $$
  \mbox{\rm deg}\,f \;
    = \;  \left| \begin{array}{cc}
                      k_1 & l_1 \\ m_1 & n_1
                 \end{array} \right|\,
        +\, \cdots\,
        + \, \left| \begin{array}{cc}
                      k_g & l_g \\ m_g & n_g
                 \end{array} \right|\,.
 $$
} 

\bigskip

\noindent
{\it Proof.} We shall not distinguish $a_i$, $b_i$ with their
corresponding simple loops in $\Sigma$. Under homotopy of
loops, one may assume that $a_i$, $b_i$ are loops at a $p_0$ in
$\Sigma$ and that the complement of the loops is a $4g$-gon $\Delta^2$,
whose boundary is given by the edge-loop $\Pi_{i=1}^g [a_i,b_i]$. Let
${\Bbb T}^2={\Bbb R}^2/({\Bbb Z}\oplus{\Bbb Z})$ with respect to the
lifting of the given basis for ${\Bbb T}^2$. Then the restriction
$f|_{\Delta^2}$ can be lifted to a map $\tilde{f}$ from $\Delta^2$
to ${\Bbb R}^2$. With an appropriate introduction of coordinates and
some abuse of notations, let $dx$, $dy$ be the $1$-forms on ${\Bbb T}$
dual to the given basis. We shall denote their lifting to the covering
${\Bbb R}^2$ also by $dx$ and $dy$. Then
$$
 \mbox{\rm deg}\,f \;
  = \; \int_{\Sigma}\,f^{\ast}(dx\wedge dy)\;
  = \; \int_{\Delta^2}\, \tilde{f}^{\ast}(dx\wedge dy)\;
  = \; \frac{1}{2}\,\int_{\partial \Delta^2}\,
                    \tilde{f}^{\ast}(x\,dy\, -\, y\, dx)\,.
$$
As indicated in {\sc Figure 3.2}, the loop
$\tilde{f}\circ\partial\Delta^2$ sweeps out $g$ curvy parallelograms,
with one for each $[a_i,b_i]$. The formula then follows. This
concludes the lemma.

\begin{figure}[htbp]
\setcaption{{\sc Figure 3.2.} The image of $\partial\,\Delta^2$ in
    the covering space ${\Bbb R}^2$ of ${\Bbb T}^2$ under
    $\tilde{f}$ sweeps out $g$ curvy parallelograms. Three of them
    are shown.  }
\centerline{\psfig{figure=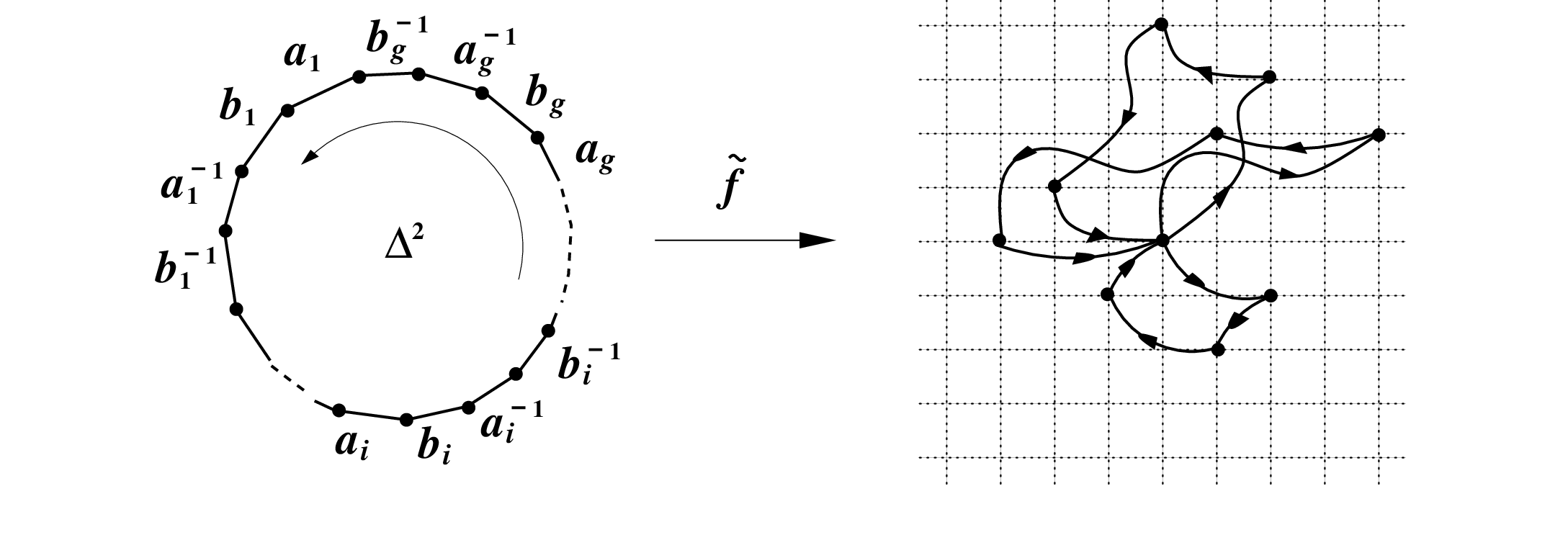,width=13cm,caption=}}
\end{figure}

\noindent \hspace{14cm} $\Box$

\bigskip

\begin{flushleft}
{\bf The tautological map: (2) When $M^3$ is closed.}
\end{flushleft}
In the following discussion, we shall not distinguish the map $f$
from $M^3$ into $N$ and its projection into
${\Bbb T}^2\times\mbox{\rm K3}$.

Recall from Remark 2.8 that up to homotopy any two representatives for
a $2$-homotopy class $[f]_2$ from $M^3$ to $N$ differ by an
amalgamation with a map $g$ from $S^3$ to $N$. Observe also that the
Hurewicz homomorphism from $\pi_3({\Bbb T}^2\times\mbox{\rm K3})$ into
$H_3({\Bbb T}^2\times\mbox{\rm K3};{\Bbb Z})$ is a zero map since
$\pi_3({\Bbb T}^2\times\mbox{\rm K3})=\pi_3(\mbox{\rm K3})$ and
$H_3(\mbox{\rm K3};{\Bbb Z})=0$. Consequently,
$$
 \langle f+g \rangle \;=\; \langle f\rangle\, +\, \langle g\rangle
    \;=\; \langle f\rangle\,.
$$
and one has

\bigskip

\noindent
{\bf Lemma 3.2.} {\it The homology class $\langle f\rangle$ in
 $H_3({\Bbb T}\times\mbox{\rm K3};{\Bbb Z})$ depends only on the
 $2$-homotopy class $[f]_2$ of $f$.
} 

\bigskip

Let $[M^3,N]_2$ be the set of $2$-homotopy classes of maps from $M^3$
to $N$. Recall from Sec.\ 2 that
$$
 \begin{array}{ccl}
   [M^3,N]_2  & =  & H^1(M^3;\pi_1({\Bbb T}^2))
                \times H^2(M^3;\pi_2(\mbox{\rm K3})) \\[1ex]
      & = &  (\oplus_2{\Bbb Z})^{b_1}
             \times \left( (\oplus_{22}\,{\Bbb Z})^{b_2}
              \,\mbox{$\bigoplus$}
            \oplus_{22}\mbox{\it Tor}\,H_1(M^3;{\Bbb Z})\right)
 \end{array}
$$
with respect to a basis $(S^1_1, S^1_2)$ for
$\pi_1({\Bbb T}^2)=H_1({\Bbb T}^2;{\Bbb Z})$ and the basis
$(S^2_1,\cdots,S^2_{22})$ for
$\pi_2(\mbox{\rm K3})=H_2(\mbox{\rm K3};{\Bbb Z})$ from the component
$2$-spheres in $\mbox{\rm K3}=(\vee_{22}\,S^2)\cup_{h_0} e^4$.
By K\"{u}nneth formula,
$$
 H_3(N;{\Bbb Z})\;
  =\; H_3({\Bbb T}^2\times \mbox{\rm K3}\,;{\Bbb Z})\;
  =\; H_1({\Bbb T}^2;{\Bbb Z})
                 \otimes H_2(\mbox{\rm K3}\,;{\Bbb Z})
$$
is generated by the $44$ classes $\langle S^1_i\times S^2_j\rangle$,
$i=1,2$ and $j=1,\cdots, 22$. With respect to this tensor product
structure and the basis, the intersection form for
$H_3({\Bbb T}^2\times\mbox{\rm K3};{\Bbb Z})$ can be expressed in the
following block form
$$
 \left(\begin{array}{cc}
    \mbox{\rm O}_{22\times 22} & 3\,\left(\begin{array}{cc}
                0 & 1 \\ 1 & 0
                 \end{array} \right) \oplus 2\,(-E_8) \\
    3\,\left(\begin{array}{cc}
                0 & 1 \\ 1 & 0
                 \end{array} \right) \oplus 2\,(-E_8)
                               & \mbox{\rm O}_{22\times 22}
       \end{array} \right)\:,
$$
where $\mbox{\rm O}_{22\times 22}$ is the $22\times 22$ zero matrix.
This matrix is non-degenerate, whose inverse is given by
$$
 \left(\begin{array}{cc}
    \mbox{\rm O}_{22\times 22} & 3\,\left(\begin{array}{cc}
                0 & 1 \\ 1 & 0
                 \end{array} \right) \oplus 2\,(-E_8^{-1}) \\
    3\,\left(\begin{array}{cc}
                0 & 1 \\ 1 & 0
                 \end{array} \right) \oplus 2\,(-E_8^{-1})
                               & \mbox{\rm O}_{22\times 22}
       \end{array} \right)\:,
$$
where
$$
 E_8^{-1}\;=\;
 \left(  \mbox{\small
 $\begin{array}{rrrrrrrr}
   4  &    5  &    7  &   10  &    8  &    6  &    4  & \; 2  \\
   5  &    8  &   10  &   15  &   12  &    9  &    6  & \; 3  \\
   7  &   10  &   14  &   20  &   16  &   12  &    8  & \; 4  \\
  10  &   15  &   20  &   30  &   24  &   18  &   12  & \; 6  \\
   8  &   12  &   16  &   24  &   20  &   15  &   10  & \; 5  \\
   6  &    9  &   12  &   18  &   15  &   12  &    8  & \; 4  \\
   4  &    6  &    8  &   12  &   10  &    8  &    6  & \; 3  \\
   2  &    3  &    4  &    6  &    5  &    4  &    3  & \; 2
 \end{array}$ }
 \right)\,.
$$
Hence one can identify $\langle f\rangle$ by its intersection
numbers with the $3$-cycles in the basis.

To understand these numbers, let us fix a Heegaard splitting
$M^3=H_g\cup_{\Sigma_g}H^{\prime}_g$ and recall the
associated complex $M^3=X^2\cup_h\,e^3$. Define a
{\it characteristic system of solid tori} in $M^3$ with respect to
the splitting to be a collection of disjoint embedded solid tori
$(S^1\times D^2)_i$, $(S^1\times D^2)_{i^{\prime}}$,
$i=1, \ldots, g$, such that
 (1) $(S^1\times D^2)_i$ (resp.\ $(S^1\times D^2)_{i^{\prime}}$)
     is contained in the interior of $H_g$ (resp.\ $H^{\prime}_g$),
     and
 (2) the core $S^1\times\{0\}$ of $(S^1\times D^2)_i$
     (resp.\ $(S^1\times D^2)_{i^{\prime}}$) intersects $e^2_i$
     (resp.\ $e^{2\,\prime}_i$) exactly once and is disjoint from
     all other $e^2_j$ (resp.\ $e^{2\,\prime}_j$)
({\sc Figure 3.3}). It follows that the collection of the core loops of
the $g$ tori in $H_g$ form a generating set for $H_1(M^3;{\Bbb Z})$; and
similarly for those in $H^{\prime}_g$.
\begin{figure}[htbp]
\setcaption{{\sc Figure 3.3.} A characteristic system of solid tori
    in $M^3$ with respect to a Heegaard splitting
    $M^3=H_g\cup_{\Sigma_g}H^{\prime}_g$. Only the part in $H_g$ are
    shown.   }
\centerline{\psfig{figure=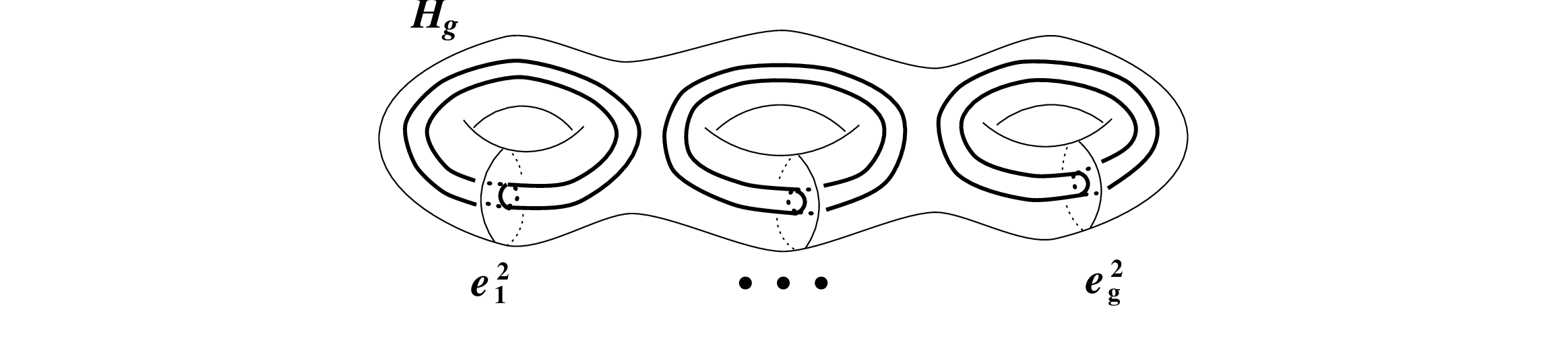,width=13cm,caption=}}
\end{figure}

Let $\mbox{\it pr}_1$, $\mbox{\it pr}_2$ be the projection map from
${\Bbb T}^2\times\mbox{\rm K3}$ into ${\Bbb T}^2$ and K3 respectively.
In the $2$-homotopy class $[f]_2$, one can choose $f$ such that the
image of $\mbox{\it pr}_2\circ f$ lies in $\vee_{22}\,S^2$ with the
preimage $(\mbox{\it pr}_2\circ f)^{-1}(\vee_{22}\,S^2-\{\ast\})$ a
collection of disjoint embedded solid tori
$$
 \left(\mbox{$\bigcup$}_{i,j;r}\,(S^1\times D^2)_{i,j;r}\right)\;
  \bigcup \;
 \left(\mbox{$\bigcup$}_{i^{\prime},j;r^{\prime}}\,
      (S^1\times D^2)_{i^{\prime},j;r^{\prime}}\right)\;
$$
in $M^3$ that satisfy:
(1) every $(S^1\times D^2)_{i,\,\cdot\,;\,\cdot}$
(resp.\ $(S^1\times D^2)_{i^{\prime},\,\cdot\,;\,\cdot}$) is isotopic
to $(S^1\times D^2)_i$ (resp.\ $(S^1\times D^2)_{i^{\prime}}$); and
(2) every $(S^1\times D^2)_{\,\cdot\,,j;\,\cdot\,}$ is mapped to
$S^2_j-\{\ast\}$ under $\mbox{\it pr}_2\circ f$.
Let $\alpha_i$ be the Poincar\'{e} dual of $S^1_i$ in ${\Bbb T}^2$,
$\beta_j$ be the Poincar\'{e} dual of $S^2_j$ in K3. Then, with some
abuse of notations, the intersection number of $\langle f\rangle$
with $\langle S^1_{i_0}\times S^2_{j_0}\rangle$ is given by
\begin{eqnarray*}
\lefteqn{
 \langle f \rangle\,\cdot\,\langle S^1_{i_0}\times S^2_{j_0}\rangle\;
 =\; \int_{\cup_{i,j;r}(S^1\times D^2)_{i,j;r}}
                      f^{\ast}(\alpha_{i_0}\wedge \beta_{j_0}) }\\
 & & \hspace{15em}
   +\; \int_{\cup_{i^{\prime},j;r^{\prime}}
       (S^1\times D^2)_{i^{\prime},j;r^{\prime}}}
                      f^{\ast}(\alpha_{i_0}\wedge \beta_{j_0})  \\
 & & \hspace{2em}
   =\; \sum_{i,j;r}\,\left(\int_{\gamma_{\,i,j,r}} \alpha_{i_0}
                       \int_{D^2_{i,j;r}} \beta_{j_0} \right)\,
   +\, \sum_{i^{\prime},j;r^{\prime}}\,
     \left(\int_{\gamma_{\,i^{\prime}j;r^{\prime}}} \alpha_{i_0}
       \int_{D^2_{i^{\prime},j;r^{\prime}}} \beta_{j_0} \right)\,,
\end{eqnarray*}
where $\gamma_{\,i,j;r}$ is the core $S^1\times \{0\}$ and
$D^2_{i,j;r}$ the slice $\{0\}\times D^2$ in
$(S^1\times D^2)_{i,j,r}$ and similarly for
$\gamma_{\,i^{\prime},j;r^{\prime}}$ and
$D^2_{i^{\prime},j;r^{\prime}}$. But this integral is exactly
the intersection number of the class
$$
 \Psi_f\; = \;\sum_{i,j;r}\,
   \langle (g^{(1)}\times g^{(2)})_{i,j;r}\rangle \;
  +\; \sum_{i^{\prime},j;r^{\prime}}\,
   \langle (g^{(1)}\times g^{(2)})_{i^{\prime},j;r^{\prime}}\rangle
$$
with $\langle S^1_{i_0}\times S^2_{j_0}\rangle$, where
$(g^{(1)}\times g^{(2)})_{\,i,j;r}$ is the map from $S^1\times S^2$
to ${\Bbb T}^2\times\mbox{\rm K3}$ given by
$$
 (g^{(1)}\times g^{(2)})_{\,i,j;r}\;
  =\; (\mbox{\it pr}_1\circ f)|_{\gamma_{\,i,j;r}}\,
     \times\, (\mbox{\it pr}_2\circ f)|_{D^2_{i,j;r}}\,;
$$
and similarly for
$(g^{(1)}\times g^{(2)})_{i^{\prime},j;r^{\prime}}$.
This shows that indeed
$$
 \langle f \rangle\; =\; \Psi_f\,.
$$

Explicitly, let $\widehat{C}_i$ (resp.\ $\widehat{C}^{\prime}_i$)
be a loop on $\Sigma_g$ homotopic to the core of $(S^1\times D^2)_i$
(resp.\ $(S^1\times D^2)_{i^{\prime}}$) in $H_g$
(resp.\ $H^{\prime}_g$). If $[\mbox{\it pr}_1\circ f]$ is represented
by a $1$-cocycle with coefficients in $\pi_1({\Bbb T}^2)$
$$
\begin{array}{cccccl}
   \eta_f & : & \{\,\widehat{C}_1, \cdots, \widehat{C}_g\,,
        \widehat{C}^{\prime}_1, \cdots, \widehat{C}^{\prime}_g\,\}
     & \longrightarrow
          & \pi_1({\Bbb T}^2)\,=\,H_1({\Bbb T}^2;{\Bbb Z}) & \\[1ex]
   & &  \widehat{C}_s   & \longmapsto
            & c_{s,1}\,\langle S^1_1\rangle\,
                 +\, c_{s,2}\,\langle S^1_2\rangle  & \\[1ex]
   & &  \widehat{C}^{\prime}_s & \longmapsto
       & c^{\prime}_{s,1}\,\langle S^1_1\rangle\,
                 +\, c^{\prime}_{s,2}\,\langle S^1_2\rangle &,
\end{array}
$$
and $[\mbox{\it pr}_2\circ f]_2$ by a
$2$-cocycle with coefficients in $\pi_2(\mbox{\rm K3})$
$$
\begin{array}{cccccl}
   \xi_f & : & \{e^2_1,\cdots, e^2_g,
              e^{2\,\prime}_1,\cdots,e^{2\,\prime}_g\}
                  & \longrightarrow  & \pi_2(\mbox{K3})\,
                             =\,H_2(\mbox{\rm K3};{\Bbb Z}) & \\[1ex]
   & &  e^2_s   & \longmapsto &  d_{s,1}\,\langle S^2_1\rangle
               + \cdots + d_{s,22}\,\langle S^2_{22}\rangle & \\[1ex]
   & &  e^{2\,\prime}_s   & \longmapsto
         &  d^{\prime}_{s,1}\,\langle S^2_1\rangle + \cdots +
                     d^{\prime}_{s,22}\,\langle S^2_{22}\rangle &,
\end{array}
$$
then $\langle f\rangle$ is the class
$$
\begin{array}{ccl}
 \langle f\rangle  & =
  & \sum_{s=1}^g\, \eta_f (\widehat{C}_s) \otimes \xi_f (e^2_s) \,
    +\, \sum_{s^{\prime}=1}^g\,
  \eta_f (\widehat{C}^{\prime}_s) \otimes \xi_f (e^{2\,\prime}_s)
                                                              \\[1ex]
 & = & \sum_{\mbox{\scriptsize
                      $\begin{array}{l}
                                i=1, 2 \\
                                j=1,\cdots,22
                       \end{array}$ }}
   \sum_{s=1}^g\, (c_{s,i}\,d_{s,j}\,
               +\, c^{\prime}_{s,i}\,d^{\prime}_{s,j})\,
                         \langle S^1_i\times S^2_j\rangle\,.
\end{array}
$$
in $H_3(N;{\Bbb Z})$.

\bigskip

\bigskip

\section{Loop orders of membrane scatterings.}

The loop order of a scattering process in quantum field and string
theory indicates the complexity of that process and serves as an
expansion parameter for the related perturbative theory.
In this last section, we discuss what happens for higher dimensional
extended objects, particularly membranes.

There are two aspects of membrane scatterings:
\begin{quote}
 \hspace{-2.1em} (1) {\it The Hamiltonian aspect}:
  Evolution of membranes and their interactions at various instances
  are emphasized. Hence the Cerf-Morse-Smale theory on $3$-manifolds
  plays the key.

 \hspace{-2.1em} (2) {\it The Lagrangian aspect}:
  The membrane world-volume is treated as a whole. Hence the topology
  of $3$-manifolds plays the key.
\end{quote}
Let us thus take a look at both aspects. Physicists are referred to
[Ce], [D-F-N], [He], [Ja], [M-B], [M-T], [Sco], [Sta], [Th1] and [Th2]
for a survey and details of miscellaneous mathematics used in this
section.

\bigskip

\begin{flushleft}
{\bf The Hamiltonian aspect.}
\end{flushleft}
Let us begin with a list of essential objects that describe a
membrane scattering and their mathematical equivalent in
Cerf-Morse-Smale theory.

\vspace{.8cm}

\centerline{\small
\begin{tabular}{lll|lll}
    && \hspace{2em}
    \begin{minipage}[t]{5.5cm}\rule{0ex}{3.5ex}
    \hspace{-.6ex}{\it Membrane scattering}:
    \rule[-2ex]{0ex}{2ex}
  \end{minipage}  & & &
      \hspace{1.5em}
   \begin{minipage}[t]{5.5cm}
    {\it Cerf-Morse-Smale theory}:
  \end{minipage}  \\ \hline
 &&\begin{minipage}[t]{6cm}
   \hspace{-10pt}$\bullet$\rule{0ex}{3.5ex}
    {\it Time-function} $\tau$ on the membrane world-volume $M^3$
    induced from the space-time where membranes are moving around.
  \end{minipage}  & & &
  \begin{minipage}[t]{6cm}
    \hspace{-10pt}$\bullet$
    {\it Morse function} $f$ on $M^3$.
  \end{minipage}  \\
 &&\begin{minipage}[t]{6cm}
    \hspace{-10pt}$\bullet$\rule{0ex}{3ex}
    Equal-time {\it slicing} of $M^3$.
  \end{minipage}  & & &
  \begin{minipage}[t]{6cm}
     \hspace{-10pt}$\bullet$
    The family $f^{-1}(a)$, $a\in{\Bbb R}$.
  \end{minipage}  \\
 &&\begin{minipage}[t]{6cm}
     \hspace{-10pt}$\bullet$\rule{0ex}{3ex}
     {\it Joining} of two membranes $\Sigma^2_1$, $\Sigma^2_2$.
  \end{minipage}  & & &
  \begin{minipage}[t]{6cm}
     \hspace{-10pt}$\bullet$
     Attaching of a {\it $1$-handle} to $\Sigma^2_1\cup\Sigma^2_2$
     with one end to $\Sigma^2_1$ and the other to $\Sigma^2_2$.
  \end{minipage} \\
 &&\begin{minipage}[t]{6cm}
     \hspace{-10pt}$\bullet$\rule{0ex}{3ex}
     {\it Splitting} of a membrane $\Sigma^2$.
  \end{minipage}  & & &
  \begin{minipage}[t]{6cm}
     \hspace{-10pt}$\bullet$
     Attaching of a {\it $2$-handle} to $\Sigma^2$ along a
     {\it separating} simple loop.
     \rule[-2ex]{0ex}{1ex}
  \end{minipage} \\
 &&\begin{minipage}[t]{6cm}
     \hspace{-10pt}$\bullet$\rule{0ex}{3ex}
     {\it Mutations} of membranes from $\Sigma^2_1$ to $\Sigma^2_2$
     with different topologies.
  \end{minipage}  & & &
  \begin{minipage}[t]{6cm}
     \hspace{-10pt}$\bullet$
      Attaching of $1$-handles with both ends on $\Sigma^2_1$
      and/or $2$-handles along a {\it non-separating} simple loop in
      $\Sigma^2_1$.
     \rule[-2ex]{0ex}{1ex}
  \end{minipage} \\
 &&\begin{minipage}[t]{6cm}
     \hspace{-10pt}$\bullet$\rule{0ex}{3ex}
     {\it Loop order $l$} and the {\it Feynman diagram} of the
     process.
  \end{minipage}  & & &
  \begin{minipage}[t]{6cm}
     \hspace{-10pt}$\bullet$
     {\it Loop number $l$} of the graph $\Gamma$ obtained by pinching
     every connected component of $f^{-1}(a)$ into a point. Note that
     $l=1-\chi(\Gamma)$, where $\chi(\Gamma)$ is the Euler
     characteristic of $\Gamma$.
     \rule[-2ex]{0ex}{1ex}
  \end{minipage} \\
 &&\begin{minipage}[t]{6cm}
     \hspace{-10pt}$\bullet$\rule{0ex}{3ex}
     Relations between different equal-time slicings.
  \end{minipage}  & & &
  \begin{minipage}[t]{6cm}
     \hspace{-10pt}$\bullet$
     Cerf's theory.
     \rule[-2ex]{0ex}{1ex}
  \end{minipage} \\
\end{tabular}
} 

\vspace{1cm}

\noindent
(Cf.\ {\sc Figure 4.1.})

\begin{figure}[htbp]
\setcaption{{\sc Figure 4.1.} The relation among membrane interaction,
    equal-time slicing, and attaching of handles are indicated. Notice
    that, at $3$-dimensions, attaching of a $1$-handle turns into
    attaching of a $2$-handle when the time direction is reversed.   }
\centerline{\psfig{figure=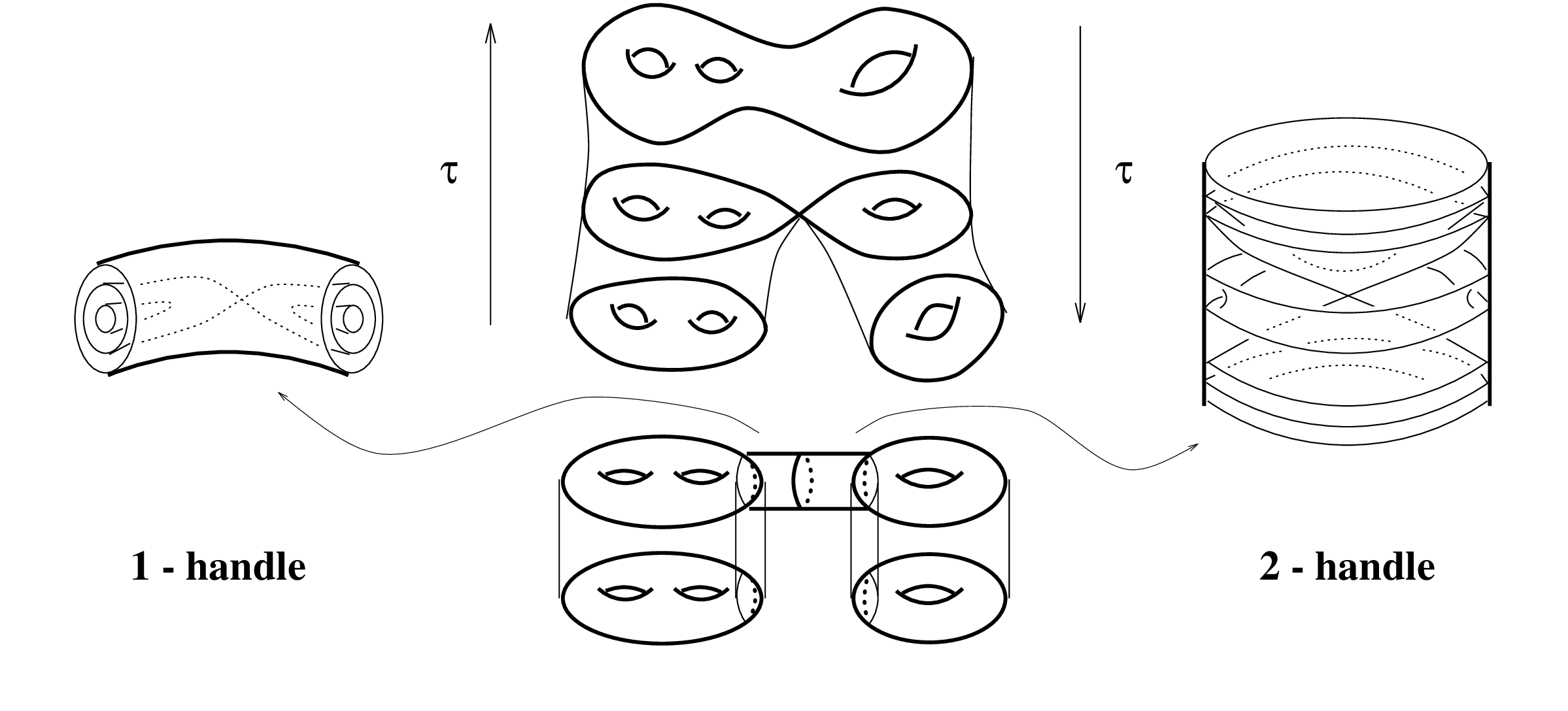,width=13cm,caption=}}
\end{figure}

This list describes well the Hamiltonian picture of string
scatterings if one replaces "membrane" by "string", "$1$-" and
"$2$-handle" at $3$-dimensions by "$1$-handle" at $2$-dimensions,
etc., and recalls the Mandelstam diagrams for string world-sheets.
However there are some new features that are not in common.

\begin{quote}
 \hspace{-2em} (1)
  While strings do not have enough room for mutations, higher
  dimensional extended objects, e.g.\ membranes, do. In the latter
  case, due to $(1,2)$-handle pair-creations and handle slidings,
  Feynman diagrams associated to a given $M^3$ can be created, whose
  number of $2$-valent vertices is greater than any given number.

 \hspace{-2em} (2)
  While the loop order $l$ thus described is well-defined for string
  world-sheets, it does depend on the equal-time slicing of the
  world-volume for higher dimensional extended objects, as indicated by
  the following example for membranes. In the latter case,
  it ranges from $0$ (by handle-slidings so that all $1$-handles come
  prior to $2$-handles or equivalently by considering a Morse-Smale
  function) to the maximal rank of the free quotient of
  $\pi_1(M^3)$.\footnote{
   I would like to thank Prof.\ William Thurston for pointing
   this out to me and providing the example.}
  Hence, for example, every compact orientable $M^3$ contributes to
  some tree-level scattering of membranes.
\end{quote}

\bigskip

\noindent
{\bf Example 4.1.} Consider $M^3=S^2\times S^1-(B^3_-\cup B^3_+)$,
which contributes to the membrane propagator from an incoming
$S^2_-=\partial B^3_-$ to an outgoing $S^2_+=\partial B^3_+$. The
two different slicings of $M^3$ as indicated in {\sc Figure 4.2}
lead to Feynman diagrams of different loop orders. This is a general
phenomenon and can be explained via handle slidings. In (a), only
mutations of $S^2$ are involved while, in (b), two Yukawa interactions
are involved. Notice that
$\pi_1(S^2\times S^1-(B^3_-\cup B^3_+))={\Bbb Z}$ has rank $1$; hence
this $M^3$ contributes only to tree and $1$-loop propagators from
$S^2_-$ to $S^2_+$.

\begin{figure}[htbp]
\setcaption{{\sc Figure 4.2.} Two different slicings of
    $S^2\times S^1-(B^3_-\cup B^3_+)$ that lead to Feynman diagrams
    of different loop orders. }
\centerline{\psfig{figure=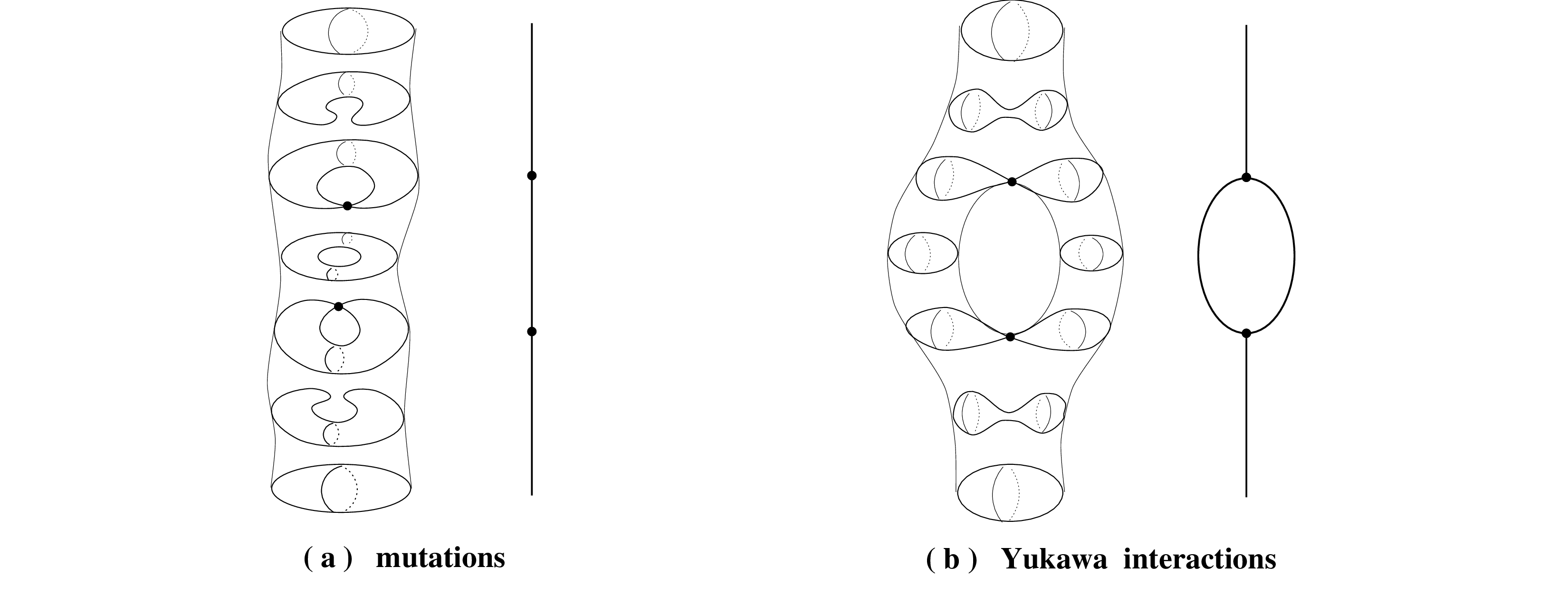,width=13cm,caption=}}
\end{figure}

\bigskip

\begin{flushleft}
{\bf The Lagrangian aspect.}
\end{flushleft}
When the membrane world-volume $M^3$ is taken as a whole,
two important elements for understanding its manifold structure are
the {\it almost canonical decompositions} of $M^3$, following
Kneser-Milnor-Waldhausen-Johannson-Jaco-Shalen (in historical order),
and the {\it Thurston's geometrization program}, indicated in
{\sc Figure 4.3} outlined from [Sco]. Though some last details remain
conjectured or unpublished, the program depicts, among other things, a
build-in complexity of a compact $3$-manifold from its very own
topology.

\begin{figure}[htbp]
\setcaption{{\sc Figure 4.3.} Decompositions and Thurston's
    geometrization program for compact $3$-manifolds. The interior of
    the compact $3$-manifold in every end of the flow chart admits a
    complete Riemannian structure locally modelled on one of the eight
    geometries: $S^3$, ${\Bbb E}^3$, ${\Bbb H}^3$, $S^2\times{\Bbb E}$,
    ${\Bbb H}^2\times{\Bbb E}$, $\mbox{\rm Nil}$,
    $\widetilde{\mbox{\it SL}\,(2;{\Bbb R})}$, and $\mbox{\rm Sol}$.
    (One may call this "{\it Thurston's "eight-fold way"}".)    }
\centerline{\psfig{figure=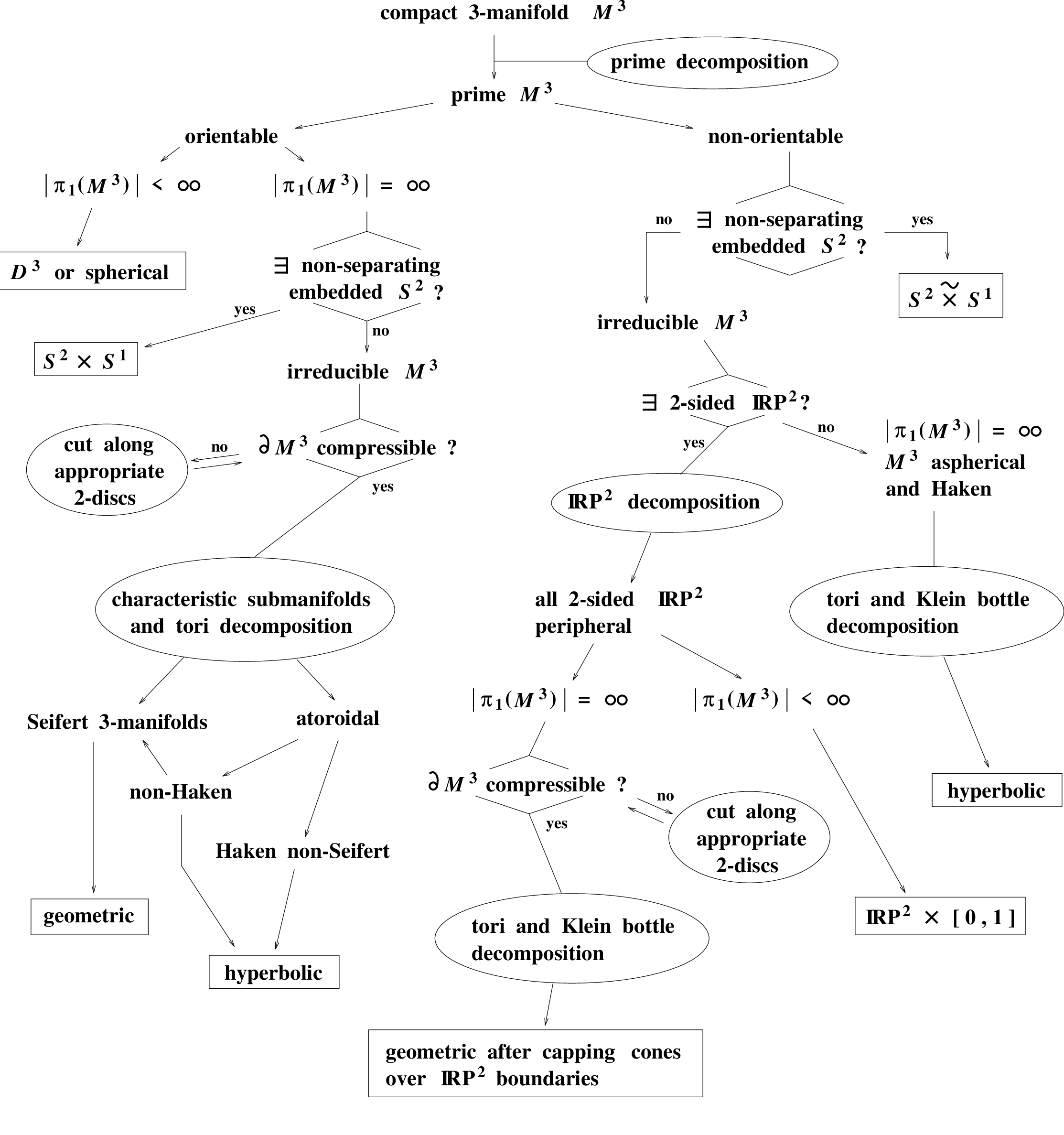,width=13cm,caption=}}
\end{figure}

As already mentioned in [Su], there are a family of graphs associated to
$M^3$ following K-M-W-J-J-S-T, as illustrated in {\sc Figure 4.4} for
$M^3$ orientable. Complexity of $M^3$, as a membrane world-volume, can
then be measured in terms of the loop number of these graphs and the
complexity of the geometric pieces appearing as vertices of the graph.
(For those pieces that are hyperbolic with finite volume, their
hyperbolic volume may serve as a measure of complexity due to the
following facts ([B-P], [Thu1]): (1) The set of volumes of (complete)
hyperbolic $3$-manifolds is well-ordered; and (2) the volume is a
finite-to-one function of (complete) hyperbolic manifolds.)

\begin{figure}[htbp]
\setcaption{{\sc Figure 4.4.}
    An almost canonical graph associated to a compact orientable
    $3$-manifold $M^3$, following K-M-W-J-J-S-T.
    In the figure,
     $\circledcirc$ indicates an incompressible boundary component;
     $\otimes$ a compressible boundary component;
     the tree from a $\otimes$ to a collection of $\blacksquare$
       indicates the chopping of a compressible boundary till it
       becomes incompressible
       - it corresponds to a co-dimension $0$ submanifold-with-boundary
         in $M^3$ with a decomposition by $2$-handles.
    The propagator between $\circ$ in the K-M part is $S^2\times[0,1]$
     while the propagator between $\blacksquare$ in the W-J-J-S part is
     ${\Bbb T}^2\times[0,1]$.       }
\centerline{\psfig{figure=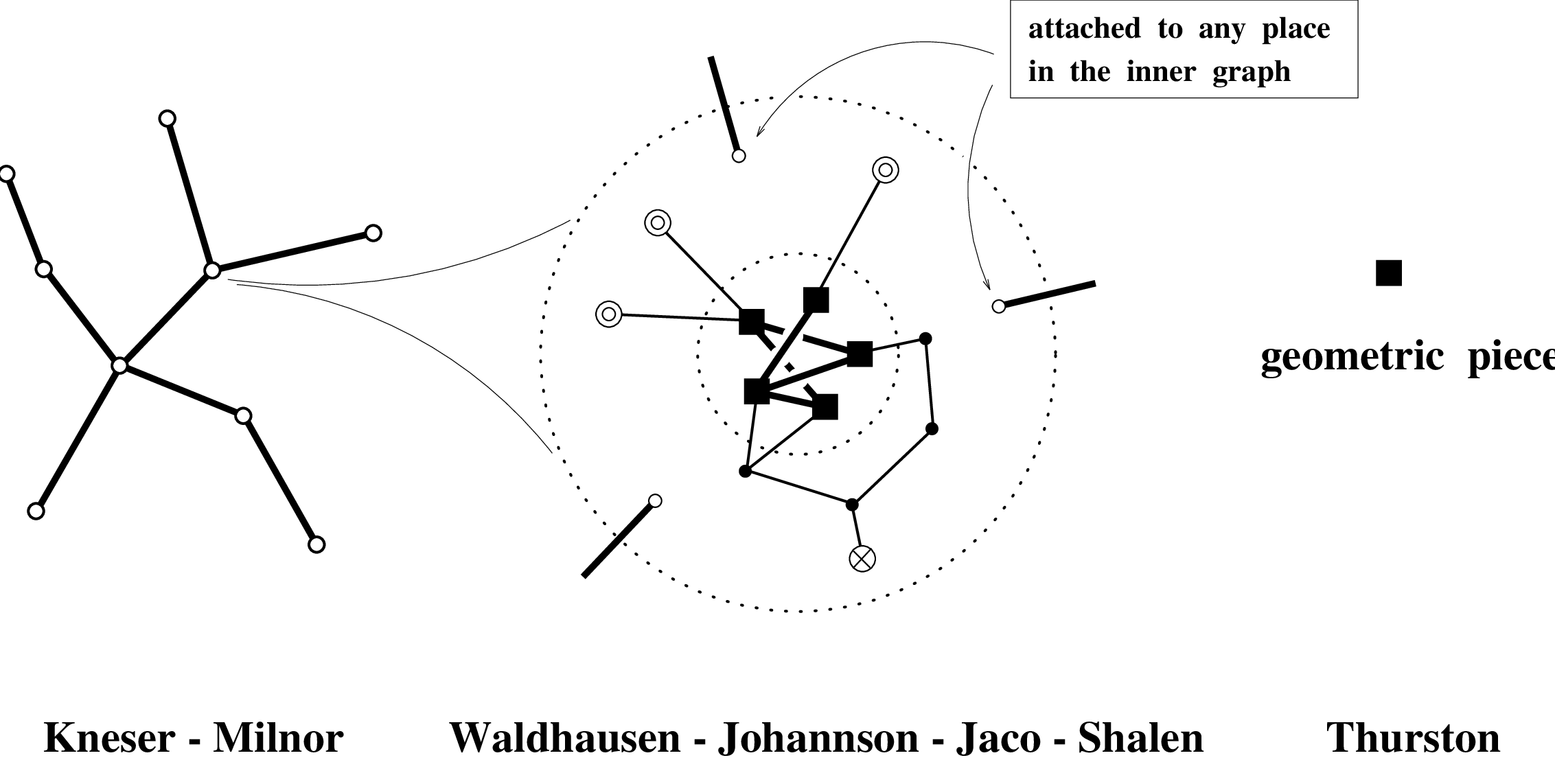,width=13cm,caption=}}
\end{figure}

On the other hand, instead of trying to fit $M^3$ into a
Feynman-diagram-like object, Milnor and Thurston explored the concept
of {\it characteristic numbers} of $3$-manifolds in [M-T], emphasizing
the natural multiplicative property under covering maps. Up to an
overall scaling, this characteristic number for hyperbolic $3$-manifolds
are their hyperbolic volumes. It is interesting to know that some
connections between hyperbolic $3$-manifolds and the physics of
membranes have been pursued by Goncharov et al.\
(cf. [Go] and some sequals). The basic starting point there is the
assumption that hyperbolic membrane world-volumes are the main
contributors to the membrane partition function in the Minkowskian
space-time. If it really turns out that these are the only
$3$-manifolds of physical significance in M-theory, then by coupling
the volume-form of a hyperbolic $M^3$ with a dilaton field - exactly
like in string theory -, they can be related to some natural scales in
M-theory (cf.\ [Sh1 - 2]).\footnote{
 I would like to thank Prof.\ Emil Martinec for a discussion
 and Prof.\ Stephen Shenker for his talk at TASI-96, that draw my
 attention to this. }
In view of this, the right generalization of loop numbers for pointlike
objects and strings to membranes may indeed be the Milnor-Thurston's
characteristic numbers. If so, one would like to know {\it
if there could be any Feynman diagram (and hence membrane scattering
process) naturally associated to this number}.

\bigskip

\noindent {\it Remark 4.2.}
Notice that there can be infinitely many (though only countable)
topologically different $M^3$ that share the same Betti numbers
$(b_1,b_2)$ (cf.\ Appendix). Thus they are not efficient in serving
as the complexity or the loop number for $M^3$.

\bigskip

The two aspects of membrane scatterings and facts from
$3$-manifolds indicate that the nature of membrane particles can be
very different from that of pointlike or stringy particles. We
conclude with the hope that future collaborations of $3$-dimensional
geometers and string theorists can unveil this secret in M-theory.

\vspace{1cm}

\begin{flushleft}
{\large\bf Appendix:} \
   \parbox[t]{12cm}{\large\bf
         A light exposition on homologies of 3-manifolds.}
\end{flushleft}

\noindent
Since $[M^3,N]$ depends quite on $H_{\ast}(M^3;{\Bbb Z})$, let us give
a light exposition on the latter for completeness. We assume that $M^3$
is compact orientable.

\bigskip

\begin{flushleft}
{\bf General facts.}
\end{flushleft}
\begin{quote}
  \hspace{-2em} (1)
   $H_2(M^3;{\Bbb Z})$ is free abelian while $H_1(M^3;{\Bbb Z})$, being
   the abelianization of $\pi_1(M^3)$, can has torsion elements.
   Two basic tools for computing $\pi_1(M^3)$ are the van Kampen's
   theorem and the homotopy sequence of a fibration ([Vi] and [Wh]).

  \hspace{-2em} (2) ([Mi2] and [Sta].)
   $\pi_1(M^3_1\,\sharp\,M^3_2)\;=\;\pi_1(M^3_1)\,\ast\,\pi_1(M^3_2)$
   and
   $$
    H_i(M^3_1\,\sharp\,M^3_2;{\Bbb Z})\,=\,
     H_i(M^3_1:{\Bbb Z})\oplus H_i(M^3_2;{\Bbb Z}),
     \hspace{1em} \mbox{for $i=1$, $2$.}
   $$

  \hspace{-2em} (3)
   For $M^3$ closed, $\chi(M^3)=0$ -
   in fact $T_{\ast}M^3$ is trivial ([Ki]) -. Hence $b_1=b_2$. In
   general, by considering the double of the $3$-manifold, one has
   $b_2-b_1=\frac{1}{2}\chi(\partial M^3)$.
\end{quote}

\bigskip

\begin{flushleft}
{\bf Examples.}
\end{flushleft}
{\bf Example A.1.} {\it Homology $3$-spheres}: ([Ro] and [Sti].)
They have trivial homologies $H_1(M^3;{\Bbb Z})=H_2(M^3;{\Bbb Z})=0$
but possibly non-trivail $\pi_1(M^3)$ ([Ro] and [Sti]). Notably the
Poincar\'{e}'s homology $3$-sphere is the only known example that has
finite $\pi_1$, of order $120$.

\bigskip

\noindent
{\bf Example A.2.} {\it Lens space $L(p,q)$}: ([Ro] and [Sti].)
They are constructed by identifying the boundary $S^2$ of a $3$-ball
$B^3$ by the map
$$
 (\theta,\phi)\; \longmapsto \; (-\theta,\phi+2\pi\cdot\frac{q}{p})\,,
$$
in terms of latitude and longitude. $L(1,q)=S^3$ has $H_1=H_2=0$.
$L(0,1)=S^2\times S^1$ has $H_1=H_2={\Bbb Z}$. For all other $L(p,q)$
with $0<q<p$ and $p$, $q$ relatively prime, $\pi_1={\Bbb Z}_p$ and
hence $H_1={\Bbb Z}_p$ and $H_2=0$.

\bigskip

\noindent
{\bf Example A.3.} {\it Knot complements in $S^3$}: ([Ro].)
They all have $H_1=H_2={\Bbb Z}$ and $\pi_i=0$, for $i\geq 2$.

\bigskip

The following two kinds of $3$-manifolds are particularly akin to
surfaces: {\it Seifert $3$-manifolds} and {\it mapping tori} of a
compact surface. In view of the aspect of membranes as excitations of
strings, they may play special roles in M-theory.

\bigskip

\noindent
{\bf Example A.4.} {\it Seifert $3$-manifolds}: They have been studied
by Seifert [Se]. Their role in understanding the general structure of
$3$-manifolds is studied in [Ja], [Jo], and [J-S]; and their geometric
structures are discussed in detail in [Sco].

In terms of the language of bundles, a Seifert $3$-manifold $M^3$ can
be defined as the total space of an $S^1$-bundle $\eta$ over a
$2$-orbifold $Q$. For $M^3$ compact and oriented, they are all
irreducible except $S^2\times S^1$ and are determined by the following
invariants:
$$
 e(\eta)\,, \hspace{1em} \chi(Q)\,, \hspace{1em}
  (p_i,q_i)\,,\; i=1, \cdots,\, r\,,
$$
where $e(\eta)\in{\Bbb Z}$ is the Euler number of the bundle,
$\chi(Q)\in{\Bbb Q}$ the Euler characteristic of the orbifold,
$r$ the number of exceptional fibers, and $(p_i,q_i)$, $p_i$ and $q_i$
relatively prime, indicates that the exceptional fiber labelled $i$ has
a fibered solid torus neighborhood $T(p_i,q_i)$ obtained by taking the
oriented cylinder $D^2\times[0,1]$, foliated by $[0,1]$, and gluing
$D^2\times\{1\}$ to $D^2\times\{0\}$ with a twist by $2\pi\frac{q}{p}$.

For $M^3$ closed oriented, $\pi_1(M^3)$ has the following presentation
[Se]:
\begin{quote}
 \hspace{-1.6em} (i) for $Q$ orientable,
 $$
 \begin{array}{c}
  \langle A_1,\, B_1,\, \cdots,\, A_g,\, B_g,\,
          Q_0,\, Q_1,\, \cdots,\, Q_r,\, H\,|  \hspace{3cm}\\[1ex]
     A_i H A_i^{-1}\,=\,H^{\epsilon_i}\,,\;
       B_i H B_i^{-1}\,=\,H^{\epsilon^{\prime}_i}\,,
         \hspace{1em} (i=1,\,\cdots,\,g;\,
           \epsilon_i,\,\epsilon^{\prime}_i\,=\,\pm 1)\,,\\[1ex]
    Q_0 Q_1\,\cdots\,Q_r\, =\, \prod_{i=1}^g\,[A_i,B_i]\,,\\[1ex]
    Q_j H Q_j^{-1}\,
         =\, H \hspace{1em} (j=0,\, 1,\, \cdots,\, r)\,, \\[1ex]
    Q_0 H^b\, =\, Q_1^{p_1} H^{\hat{q}_1}\,
    =\, \cdots\, =\, Q_r^{p_r} H^{\hat{q}_r}\,  =\, 1\,\rangle\,;
 \end{array}
 $$

 \hspace{-2em} (ii) for $Q$ non-orientable,
 $$
 \begin{array}{c}
  \langle A_1,\, \cdots,\, A_g,\,
          Q_0,\, Q_1,\, \cdots,\, Q_r,\, H\,| \hspace{3cm}\\[1ex]
     A_i H A_i^{-1}\,=\,H^{\epsilon_i}\,,\;
         \hspace{1em} (i=1,\,\cdots,\,g;\,
           \epsilon_i,\,\epsilon^{\prime}_i\,=\,\pm 1)\,,\\[1ex]
    Q_0 Q_1\,\cdots\,Q_r\, =\, \prod_{i=1}^g\,A_i^2\,,\\[1ex]
    Q_j H Q_j^{-1}\,
         =\, H \hspace{1em} (j=0,\, 1,\, \cdots,\, r)\,, \\[1ex]
    Q_0 H^b\, =\, Q_1^{p_1} H^{\hat{q}_1}\,
    =\, \cdots\, =\, Q_r^{p_r} H^{\hat{q}_r}\, =\, 1\, \rangle\,,
 \end{array}
 $$
\end{quote}
where
$g=1-\frac{1}{2}\chi(Q)-\frac{1}{2}\sum_{i=1}^r(1-\frac{1}{p_i})$
for $Q$ orientable, $=2-\chi(Q)-\sum_{i=1}^r(1-\frac{1}{p_i})$ for
$Q$ non-orientable, is the genus of $Q$ (i.e.\ the number of handles
for $Q$ orientable or the number of crosscaps for $Q$ non-orientable),
$b=e(\eta)$, and $0<\hat{q}_i<p_i$ with
$q_i\hat{q}_i\equiv 1\,(\mbox{\rm mod}\;\;p_i)$. Consequently,
$H_1(M^3;{\Bbb Z})$ is isomorphic to:
\begin{quote}
 \hspace{-2em} (i') For $Q$ orientable,
 $$
  \oplus_{2g}\,{\Bbb Z}\, \mbox{$\bigoplus$}\,
   \left.\mbox{\raisebox{1ex}{
     $\oplus_{r+2}\,{\Bbb Z}$}}
    \right/ \mbox{\raisebox{-1ex}{$\Gamma$}} \;\;,
 $$
 where $\oplus_{2g}\,{\Bbb Z}$ is generated by $A_i$, $B_i$,
 $\oplus_{r+2}\,{\Bbb Z}$ by $H$, $Q_j$, and $\Gamma$ is the
 lattice generated by
 $$
 \begin{array}{c}
  \hspace{3em} \epsilon\,H\,, \hspace{1em} \mbox{where
   $\epsilon\,=\,\mbox{\it max}\,_i\{1-\epsilon_i\}\,
                         =\, \mbox{$0$ or $2$}$}\,, \\[1ex]
   Q_0 + Q_1 +\,\cdots\, + Q_r\,, \\[1ex]
   Q_0 + b\,H\,,\;\;
    p_1 Q_1 + \hat{q}_1 H\,,\;\;
     \cdots\,,\;\; p_r Q_r + \hat{q}_r H\,.
 \end{array}
 $$
\end{quote}
The rank of the above $r+3$ vectors is
$$
\begin{array}{ll}
 r+1  & \mbox{if both $\epsilon$ and
       $b p_1 \ldots p_r + \sum_{i=1}^{r-1} \hat{q}_i p_1 \ldots
        \nearrow\hspace{-2.4ex}p_i \ldots p_{r-1}$ are $0$,
        where}\\[1ex]
  & \mbox{$p_1 \ldots \nearrow\hspace{-2.4ex}p_i \ldots p_{r-1}$
    is the product $p_1 \ldots p_{r-1}$ with $p_i$ deleted,}\\[1ex]
 r+2  & \mbox{otherwise.}
\end{array}
$$

\begin{quote}
 \hspace{-2.2em} (ii') For $Q$ non-orientable,
 $$
   \left.\mbox{\raisebox{1ex}{
     $\oplus_{g+r+2}\,{\Bbb Z}$}}
    \right/ \mbox{\raisebox{-1ex}{$\Gamma$}} \;\;,
 $$
 where $\oplus_{g+r+2}\,{\Bbb Z}$ is generated by $A_i$, $H$ and $Q_j$,
 and $\Gamma$ is the lattice generated by
 $$
 \begin{array}{c}
  \hspace{3em} \epsilon\,H\,, \hspace{1em} \mbox{where
   $\epsilon\,=\,\mbox{\it max}\,_i\{1-\epsilon_i\}\,
                                =\, \mbox{$0$ or $2$}$}\,, \\[1ex]
   Q_0 + Q_1 +\,\cdots\, + Q_r - 2A_1 -\,\cdots\, - 2A_g\,, \\[1ex]
   Q_0 + b\,H\,,\;\;
    p_1 Q_1 + \hat{q}_1 H\,,\;\;
     \cdots\,,\;\; p_r Q_r + \hat{q}_r H\,.
 \end{array}
 $$
\end{quote}
The rank of the above $r+3$ vectors is $r+2$ if $\epsilon=0$,
$r+3$ if $\epsilon\neq 0$.

\bigskip

\noindent
{\bf Example A.5.} {\it Mapping tori of compact surfaces}:
([C-B] and [Th5 -- 6].)
The mapping torus $\Sigma_{\phi}$ of an automorphism $\phi$ of a
compact surface $\Sigma$ is the $3$-manifold obtained by first
forming $\Sigma\times[0,1]$ and then gluing $\Sigma\times\{1\}$ to
$\Sigma\times\{0\}$ via $\phi$. Naturally, $\Sigma_{\phi}$ fibers
over $S^1$ with monodromy $\phi$. For $\Sigma$ of negative Euler
characteristic, the geometric structure of $\Sigma_{\phi}$ is
clarified by Thurston in [Th5], where it is shown that the interior
of $\Sigma_{\phi}$ either
\begin{quote}
 \hspace{-1.6em} (i)
  admits a complete ${\Bbb H}^2\times{\Bbb E}$ structure of finite
  volume, and can be described as a Seifert fibration over some
  hyperbolic $2$-orbifold,

 \hspace{-2em} (ii)
  contains an embedded non-peripheral incompressible torus, which
  splits $\Sigma_{\phi}$ into two simpler $3$-manifolds, or

 \hspace{-2.2em} (iii) (generic case)
  admits a complete hyperbolic structure of finite volume.
\end{quote}
(Cases (i) and (ii) are not mutually exclusive, but (iii) excludes
the other two.)

These three cases correspond exactly to his classification of
surface automorphisms ([C-B] and [Th6]):
\begin{quote}
 \hspace{-2em} (i')
  $\phi$ is isotopic to an automorphism of finite order,

 \hspace{-2.2em} (ii')
  $\phi$ is isotopic to a reducible automorphism, which leaves
  invariant a system of simple loops, or

 \hspace{-2.5em} (iii')
  $\phi$ is pseudo-Anosov.
\end{quote}

For homologies, let $\gamma_1$, $\cdots$, $\gamma_r$ be a set of
generators for $\pi_1(\Sigma,p_0)$ and $\sigma$ be a generator
for $\pi_1(S^1)$. Then
$$
 \pi_1(\Sigma_{\phi})\;
 =\; \left.\mbox{\raisebox{1ex}{
    $\pi_1(\Sigma,p_0)\,\ast\,\pi_1(S^1)$}}
   \right/ \mbox{\raisebox{-1ex}{$\langle\,
      \sigma\gamma_i\sigma^{-1} (\phi_{\ast}\gamma_i)^{-1};\;
                       i=1,\,\cdots,\, r \,\rangle$}} \;\;,
$$
where $\langle\,\cdots\,\rangle$ is the subgroup generated by $\cdots$,
and one identifies $\pi_1(\Sigma,p_0)$ and $\pi_1(\Sigma,\phi(p_0))$
by a fixed path connecting $p_0$ and $\phi(p_0)$. After abelianization,
one has
$$
 H_1(\Sigma_{\phi};{\Bbb Z}) \;
  =\; \left.\mbox{\raisebox{1ex}{
    $H_1(\Sigma;{\Bbb Z})\,\oplus\,H_1(S^1;{\Bbb Z})$}}
   \right/ \mbox{\raisebox{-1ex}{$\langle\,
      \gamma_i - \phi_{\ast}\gamma_i;\;
                       i=1,\,\cdots,\, r \,\rangle$}} \;\;.
$$

\bigskip

Finally, we conclude incompletely with the following broad and
beautiful subject.

\bigskip

\noindent
{\bf Example A.6.} {\it Geometic 3-manifolds}:
(E.g.\ [Sco] and [Th1 -- 6].)
Their fundamental groups are discrete subgroups of the group of
isometries of the eight model geometries. The study of the case of
hyperbolic $3$-manifolds is majorly promoted by Thurston among other
figures and is related to the study of Kleinian, Fuchsian, and
quasi-Fuchsian groups. Nice expositions are given in, e.g.\
[Sco] and [Th1 -- 3].


\vspace{1.5cm}

\footnotesize

\begin{thebibliography}{AAAAaaa}

\bibitem[As]{} P.S. Aspinwall,
 {\it An $N=2$ dual pair and a phase transition},
 {\sl Nucl. Phys.} {\bf B460} (1996), pp. 57 -- 76.

\bibitem[B-B-S]{} K. Becker, M. Becker, and A. Strominger,
 {\it Fivebranes, membranes and non-perturbative string theory},
 {\sl Nucl. Phys.} {\bf B456} (1996), pp. 130 -- 152.

\bibitem[B-P]{} R. Benedetti and C. Petronio,
 {\sl Lectures on hyperbolic geometry}, Springer-Verlag, 1992.

\bibitem[B-T]{} R. Bott and L.W. Tu,
 {\sl Differential forms in algebraic topology}, GTM 82,
 Springer-Verlag, 1982.

\bibitem[B-V-S]{} M. Bershadsky, C. Vafa, and V. Sadov,
 {\it $D$-strings on $D$-manifolds},
 {\sl Nucl. Phys.} {\bf B463} (1996), pp. 398 -- 414.

\bibitem[B- et al.]{} M. Berkooz, R.G. Leigh, J. Polchinski,
 J.H. Schwarz, N. Seiberg, and E. Witten,
 {\it Anomalies, dualities, and topology of $D=6$, $N=1$ superstring
  vacua}, {\sl Nucl. Phys.} {\bf B475} (1996), pp. 115 -- 148.

\bibitem[C-B]{} A.J. Casson and S.A. Bleiler,
 {\sl Automorphisms of surfaces after Nielsen and Thurston},
 LMSST 9, Cambridge Univ. Press, 1988.

\bibitem[Ce]{} J. Cerf,
 {\it La stratification naturelle des espaces fonctions
 diff\'{e}rentiables r\'{e}eles et le th\'{e}or\`{e}me de la
 pseudo-isotopie}, {\sl Publ. Math. I.H.E.S.} {\bf 39} (1970),
 pp. 5 -- 173.

\bibitem[Co]{} S. Coleman,
 {\it Classical lumps and their quantum descendants} in
 {\sl Aspects of symmetry}, pp. 185 -- 264, Cambridge Univ. Press,
 1985.

\bibitem[D-F-N]{} B.A. Dubrovin, A.T. Fomenko, and S.P. Novikov,
 {\sl Modern geometry - methods and applications, Part II: The
 geometry and topology of manifolds}, GTM 104;
 {\it Part III: Introduction to homology theory}, GTM 124;
 Springer-Verlag, resp.\ 1984 and 1990.

\bibitem[D-K]{} S.K. Donaldson and P.B. Kronheimer,
 {\sl The geometry of four-manifolds}, Oxford Univ. Press, 1990.

\bibitem[D-K-L]{} M.J. Duff, R.R. Khuri, and J.X. Lu,
 {\it String solitons}, {\sl Phys. Reports} {\bf 259} (1995),
 pp. 213 -- 326.

\bibitem[D-M]{} K. Dasgupta and S. Mukhi,
 {\it Orbifolds of M-theory}, {\sl Nucl. Phys.} {\bf B465} (1996),
 pp. 399 -- 412.

\bibitem[F-N]{} P.G.O. Freund and R.I. Nepomechie,
 {\it Unified geometry of antisymmetric tensor gauge fields and
  gravity}, {\sl Nucl. Phys.} {\bf B 199} (1982), pp. 482 -- 494.

\bibitem[Go]{} Yu.P. Goncharov,
 {\it Determinants of Laplacians in real line bundles over hyperbolic
  manifolds connected with quantum geometry of membranes},
 {\sl Lett. Math. Phys.} {\bf 19} (1990), pp. 73 -- 81.

\bibitem[G-H]{} P. Griffiths and J. Harris,
 {\sl Principles of algebraic geometry},
 John Wiley \& Sons, Inc., 1978.

\bibitem[G-S-W]{} M.B. Green, J.H. Schwarz, and E. Witten,
 {\sl Superstring theory}, vol.\ 1 and vol.\ 2,
 Cambridge Univ. Press, 1987.

\bibitem[Ha]{} W. Haken,
 {\it Some results on surfaces in $3$-manifolds}, in
 {\sl Studies in modern topology}, P.J. Hilton ed.,
 MAA Studies in Math., vol. 5, pp. 39 -- 98,
 Math. Asso. Amer., 1968; printed by Prentice-Hall, Inc..

\bibitem[He]{} J. Hempel,
 {\sl 3-manifolds}, Ann. Math. Studies 86, Princeton Univ. Press, 1976.

\bibitem[Hi]{} P.J. Hilton,
 {\it On the homotopy groups of the union of spheres}, {\sl J. London
  Math. Soc.} {\bf 30} (1955), pp. 154 -- 172.

\bibitem[Hir]{} M. Hirsch,
 {\it The embedding of bounding manifolds in Euclidean space},
 {\sl Ann. Math.} {\bf 74} (1961), pp. 494 -- 497.

\bibitem[Hu]{} S.-T. Hu,
 {\sl Homotopy theory}, Pure Appl. Math. vol. 8, Academic Press, 1959.

\bibitem[HA-M-S]{} C. Hog-Angeloni, W. Metzler, and A.J. Sieradski eds.,
 {\sl Two-dimensional homotopy and combinatorial group theory},
 Londen Math. Soc. Lect. Notes Ser. 197, Cambridge Univ. Press, 1993.

\bibitem[H-W1]{} P. Ho\v{r}ava and E. Witten,
 {\it Heterotic and type I string dynamics from eleven dimensions},
 {\sl Nucl. Phys.} {\bf B460} (1996), pp. 506 -- 524.

\bibitem[H-W2]{} --------,
 {\it Eleven-dimensional supergravity on a manifold with boundary},
 {\sl Nucl. Phys.} {\bf B475} (1996), pp. 94 -- 114.

\bibitem[Ja]{} W. Jaco,
 {\sl Lectures on three-manifold topology},
 Regional Conf. Ser. Math. no. 43, Amer. Math. Soc., 1980.

\bibitem[Jo]{} K. Johannson,
 {\sl Homotopy equivalence of $3$-manifolds with boundaries},
 Lect. Notes Math. no. 761, Springer-Verlag, 1979.

\bibitem[J-S]{} W. Jaco and P.B. Shalen,
 {\it A new decomposition theorem for irreducible sufficiently-large
  3-manifolds}, in {\sl Algebraic and geometric topology, Part 2},
 R.J. Milgram ed., Proc. Symp. Pure Math. vol. 32, pp. 71 -- 84,
 Amer. Math. Soc., 1978.

\bibitem[Ki]{} R.C. Kirby,
 {\sl The topology of 4-manifolds}, Lect. Notes Math. no. 1374,
 Springer-Verlag, 1989.

\bibitem[K-M]{} D. Kutasov and E.J. Martinec,
 {\it New principles for string/membrane unification},
 {\tt hep-th/9602049}.

\bibitem[K-M-O'L]{} D. Kutasov, E.J. Martinec, and M. O'Loughlin,
 {\it Vacua of M-theory and $N=2$ strings},
 {\tt hep-th/9603116}.

\bibitem[L-M]{} H. Luckock and I. Moss,
 {\it The quantum geometry of random surfaces and spinning membranes},
 {\sl Class. Quantum Grav.} {\bf 6} (1989), pp. 1993 -- 2027.

\bibitem[Ma]{} E.J. Martinec,
 {\it Geometric structures of M-theory},
 {\tt hep-th/9608017}.

\bibitem[Mi1]{} J. Milnor,
 {\it On simply connected 4-manifolds}, {\sl Symposium Internacional
  de Topolog\'{i}a Algebrica}, La Univ. Nacional Aut\'{o}noma de
 M\'{e}xico, UNESCO, 1958.

\bibitem[Mi2]{} --------,
 {\it A unique factorization theorem for 3-manifolds},
 {\sl Amer. J. Math.} {\bf 84} (1962), pp. 1 -- 7.

\bibitem[Mi3]{} --------,
 {\sl Topology from the differentiable viewpoint}, Univ. Press of
 Virginia, 1965.

\bibitem[M-B]{} J.W. Morgan and H. Bass eds.,
 {\sl The Smith conjecture}, Pure Appl. Math. vol. 112, Academic Press,
 1984.

\bibitem[M-T]{} J. Milnor and W.P. Thurston,
 {\it Characteristic numbers of 3-manifolds},
 {\sl L'Enseignement Math.} {\bf 23} (1977), pp. 249 -- 254.

\bibitem[Ne1]{} R.I. Nepomechie,
 {\it Approaches to a non-abelian antisymmetric tensor gauge field
  theory}, {\sl Nucl. Phys.} {\bf B212} (1983), pp. 301 -- 320.

\bibitem[Ne2]{} --------,
 {\it Magnetic monopoles from antisymmetric tensor gauge fields},
 {\sl Phys. Rev.} {\bf D 31} (1985), pp. 1921 -- 1924.

\bibitem[Po]{} V. Po\'{e}naru,
 {\it On the geometry of differentiable manifolds}, in
 {\sl Studies in modern topology}, P.J. Hilton ed.,
 MAA Studies in Math. vol. 5, pp. 165 -- 207,
 Math. Asso. Amer., 1968; printed by Prentice-Hall, Inc..

\bibitem[P-C-J]{} J. Polchinski, S. Chaudhuri, and C.V. Johnson,
 {\it Notes on D-branes}, {\tt hep-th/9602052}.

\bibitem[Ro]{} D. Rolfsen,
 {\sl Knots and links}, Math. Lect. Ser. 7, 2nd printing,
 Publish or Perish, Inc., 1990.

\bibitem[Sch]{} C. Schmidhuber,
 {\it $D$-brane actions}, {\sl Nucl. Phys.} {\bf B467} (1996),
 pp. 146 -- 158.

\bibitem[Sco]{} P. Scott,
 {\it The geometry of 3-manifolds}, {\sl Bull. London Math. Soc.}
 {\bf 15} (1983), pp. 401 -- 487.

\bibitem[Sh1]{} S.H. Shenker,
 {\it The strength of non-perturbative effects in string theory}, in
 {\sl Random surfaces and quantum gravity}, O. Alvarez, E. Marinari,
 and P. Windey eds., pp. 191 -- 200, Plenum Press, 1991.

\bibitem[Sh2]{} --------,
 {\it Another length scale in string theory?},
 {\tt hep-th/9509132}.

\bibitem[Se]{} H. Seifert,
 {\it Topologie dreidimensionaler gefaserte R\"{a}ume},
 {\sl Acta Math.} {\bf 60} (1933), pp. 147 -- 238.
 (English translation by W. Heil in
 {\sl A textbook of topology}, H. Seifert and W. Threlfall,
 Pure and Appl. Math. 89, Academic Press, 1980.)

\bibitem[Si]{} J. Singer,
 {\it Three-dimensional manifolds and their Heegaard diagrams},
 {\sl Trans. Amer. Math. Soc.} {\bf 35} (1933), pp. 88 -- 111.

\bibitem[Sp1]{} E.H. Spanier,
 {\it The homology of Kummer manifolds}, {\sl Proc. Amer. Math. Soc.}
 {\bf 7} (1956), pp. 155 -- 160.

\bibitem[Sp2]{} --------,
 {\sl Algebraic topology}, Springer-Verlag, 1966.

\bibitem[Sta]{} J. Stallings,
 {\sl Group theory and three-dimensional manifolds}, Yale Math.
 Monographs 4, Yale Univ. Press, 1971.

\bibitem[Sti]{} J. Stillwell,
 {\sl Classical topology and combinatorial group theory}, 2nd ed.,
 GTM 72, Springer-Verlag, 1993.

\bibitem[Su]{} A. Sugamoto,
 {\it Starting from strings to membranes}, in
 {\sl Supermembranes and physics in $2+1$ dimensions},
 M.J. Duff, C.N. Pope, and E. Sezgin eds., pp. 16 -- 28,
 World Scientific, 1990.

\bibitem[Th1]{} W.P. Thurston,
 {\sl The geometry and topology of three-manifolds},
 Princeton Lecture Notes, 1979. Also revised version 1990.

\bibitem[Th2]{} --------,
 {\it Three-dimensional manifolds, Kleinian groups and hyperbolic
 geometry}, {\sl Bull. Amer. Math. Soc.} {\bf 6} (1982),
 pp. 357 -- 381.

\bibitem[Th3]{} --------,
 {\it Hyperbolic geometry and $3$-manifolds}, in
 {\sl Low-dimensional topology}, R. Brown and T.L. Thickstun eds.,
 LMS 48, pp. 9 -- 25, Cambridge Univ. Press, 1982.

\bibitem[Th4]{} --------,
 {\it Hyperbolic structures on $3$-manifolds, I:
  Deformation of acylindrical manifolds},
 {\sl Ann. Math.} {\bf 124} (1986), pp. 203 -- 246.

\bibitem[Th5]{} --------,
 {\it Hyperbolic structures on $3$-manifolds, II:
  Surface groups and $3$-manifolds which fiber over the circle},
 Princeton preprint, 1986.

\bibitem[Th6]{} --------,
 {\it On the geometry and dynamics of diffeomorphisms of surfaces},
 {\sl Bull. Amer. Math. Soc.} {\bf 19} (1988), pp. 417 -- 431.

\bibitem[Vi]{} J.W. Vick,
 {\sl Homology theory: an introduction to algebraic topology},
 Pure Appl. Math. vol. 53, Academic Press, 1973.
 Reprinted, 2nd ed., GTM 145, Springer-Verlag, 1994.

\bibitem[Wa]{} F. Waldhausen,
 {\it Recent results on sufficiently large 3-manifolds}, in
 {\sl Algebraic and geometric topology, Part 2},
 R.J. Milgram ed., Proc. Symp. Pure Math. vol. 32, pp. 21 -- 38,
 Amer. Math. Soc., 1978.

\bibitem[Wh]{} G.W. Whitehead,
 {\sl Elements of homotopy theory}, GTM 61, Springer-Verlag, 1978.

\bibitem[Whi]{} J.H.C. Whitehead,
 {\it On simply-connected 4-dimensional polyhedra},
 {\sl Comment. Math. Helv.} {\bf 22} (1949), pp. 48 -- 92.

\bibitem[Wi1]{} E. Witten,
 {\it String theory dynamics in various dimensions},
 {\sl Nucl. Phys.} {\bf B 443} (1995), pp. 85 -- 126.

\bibitem[Wi2]{} --------,
 {\it Five-branes and M-theory on an orbifold},
 {\sl Nucl. Phys.} {\bf B463} (1996), pp. 383 -- 397.

\bibitem[Wi3]{} --------,
 {\it Phase transitions in M-theory and F-theory},
 {\sl Nucl. Phys.} {\bf B471} (1996), pp. 195 -- 216.

\bibitem[Y-Z]{} S.-T. Yau and E. Zaslow,
 {\it BPS states, string duality, and nodal curves on K3},
 {\sl Nucl. Phys.} {\bf B471} (1996), pp. 503 -- 512.

\end{thebibliography}
\end{document}